MP USER'S GUIDE  (Fourth Edition)

Richard P. Brent

Department of Computer Science

Australian National University

Box 4, Canberra, ACT 2600

Australia

Technical Report TR-CS-81-08

June 1981

CONTENTS

# 1. GENERAL DESCRIPTION OF MP

## 1.1 Introduction

MP is a multiple-precision floating-point arithmetic package.  It is almost completely machine-independent, and should run on any machine with an ANSI Standard Fortran (ANS X3.9-1966) compiler, sufficient memory, and a wordlength (for integer arithmetic) of at least 16 bits. A precompiler (Augment) which facilitates the use of the MP package is available - see Section 4 for details.

MP has been tested on various machines including a Univac 1108, a Univac 1100/82, a DEC 10, an IBM 360/50, 360/91 and 370/168, and a PDP 11/45, with various Fortran compilers.  These machines have effective integer wordlengths ranging from 16 to 48 bits.  A wordlength of 12 bits would probably be sufficient, but this has not been tested.

MP works with normalized floating-point numbers.  The base (B) and number of digits (T) are arbitrary, subject to some restrictions given below, and may be varied dynamically.

T-digit floating-point numbers are stored in integer arrays of dimension at least T+2, with the following conventions

        word 1 = sign (0, -1 or +1),
        word 2 = exponent (to base B),
        words 3 to T+2 = normalized fraction  (one base B digit per word).

Note that words 2 to T+2 are undefined if the sign is zero.  Integer arrays used for storage of T-digit floating-point numbers in the above format will be referred to as (unpacked) multiple-precision variables or (unpacked) multiple-precision numbers.  There is also a packed representation, where two base B digits are stored in each word - for details see the description of MPPACK in Section 6.

The user may ask for rounded or truncated arithmetic to be used.  With rounded arithmetic the basic arithmetic operations (addition, multiplication and division) give the correctly rounded T-digit, base B result.  With truncated arithmetic the relative error in the result is at most $B^{**}(1-T)$.  Directed roundings (useful for interval arithmetic) are also available - for details see Section 3.

Rounding options are implemented for some elementary and special functions (e.g. MPSQRT) but not yet for others.  For details see Section 6.  Most routines give at worst a result
$y = f(x)$ which could be obtained by making an $O(B^{**}(1-T))$ relative perturbation in the argument x, evaluating the function f exactly, then making an $O(B^{**}(1-T))$ relative perturbation in the result.

Exponents can lie in the range 1-M, ... , +M inclusive, where M is set by the user.  Thus, numbers from $B^{**}(-M)$ to almost $B^{**}M$ are representable.  On underflow during an arithmetic operation, the result is set to zero by subroutine MPUNFL. (This is the default action and may be modified - see Section 6.)  On overflow subroutine MPOVFL is called and execution is terminated with an error message.

Error messages are printed on logical unit LUN, where LUN is set by  the
user, and then execution is terminated by a call to subroutine MPERR.

The parameters B, T, M, LUN etc. are stored in a labelled COMMON  block,
COMMON  /MPCOM/.  A working array of sufficient size must be provided in
blank COMMON.  Details are given in Section 3.1.

1.2 Restrictions and assumptions

B (the base) must be at least 2.

T (the number of digits) must be at least 2.

M  (the  exponent range) must be greater than T and less than 1/4 of the
largest machine-representable integer.

8*B**2-1 must not exceed the largest machine-representable integer.

B**(T-1) should be at least 10**7 (this restriction does  not  apply  to
the basic arithmetic operations).

B, T etc. must be initialized by calling either MPINIT or MPSET2  before
calling any other MP routines.  For details see Section 6.

Blank COMMON rather than labelled COMMON is  used  for  working  storage
because  a  curious  restriction  in  the  ANSI (1966) Fortran standard
requires that a labelled COMMON block be declared with the  same  length
in  each  subprogram  in  which  it  is  declared, but this restriction
fortunately does not apply to blank COMMON.  Although  space  in  blank
COMMON  must  be provided for the MP routines to use as working storage,
this space need not start at the first word of  blank  COMMON  (see  the
description  of  MPSET2 in Section 6).  Thus, programs which already use
blank COMMON can make use of the MP package.

MP  variables  used  as  arguments  of subroutines need not be distinct.
Thus
      CALL MPADD (X, X, X)
and
      CALL MPEXP (X, X)

are  allowable.  However,  distinct  arrays which overlap should not be
used.  For example,  CALL MPEXP (X(1), X(2)) is not allowed if X  is  an
integer array holding a multiple-precision variable.

It is assumed that the compiler  passes  addresses  of  arrays  used  as
arguments  in  subroutine  calls  (i.e. call by reference), and does not
check for array bounds violations (either for arguments  of  subroutines
or for arrays in COMMON).  Apart from these violations, MP is written in
a portable subset of ANSI Standard Fortran (ANSI X3.9-1966).   This  has
been  checked  by the PFORT verifier.  The only machine-dependent routine
is MPUPW (which unpacks characters stored several  to  a  word).   Other
routines  which  may require trivial changes are MPINIT, MPIS and MPLARG
(see comments in Sections 6 and 9 below).  For comments on  Fortran  77,
see Section 9.4.

1.3 Note on correctness

It is clearly impossible to thoroughly test even one MP routine with all possible combinations of B, T, M etc. It is also impracticable to formally prove correctness of any nontrivial MP routines using present theorem-proving techniques. Thus, no warranty can be given that the MP package is correct. To gain confidence in it we rely on

1. Careful modular design, using the idea of 'information hiding' and the avoidance of 'tricks' and 'side-effects' as far as possible.

2. Error analysis (in the style of Wilkinson).

3. Limited testing including, where possible, comparison of results obtained by at least two independent methods (e.g. MPLN, MPLNGS and MPLNI compute logarithms by three independent methods).

Although no absolute assurance is possible, a user may be confident that a result obtained using MP is correct if

1. He uses more digits than are strictly necessary.

2. He repeats the computation using a different base B.

3. He uses two independent mathematical formulations of the problem.

It is unlikely that a bug in MP, the user's Fortran compiler, or the user's hardware, would remain undetected if steps 1-3 were followed.

MP makes minimal assumptions - primarily that INTEGER arithmetic (on numbers which are not too large) is exact.

MP does not use Fortran library routines such as SIN, ALOG, EXP, SQRT, and uses only 'A1' format for output of multiple-precision numbers.

REAL and DOUBLE PRECISION arithmetic are used only in routines for conversion of these types to and from multiple-precision etc.

1.4 History and likely future development

The first working version of MP (version 731101) was written in November 1973. Between 1973 and 1978 a considerable number of improvements and additions were made, and nonstandard Fortran constructs were eliminated. In April 1978 the Augment interface routines (see Section 4.1) were added and the first version of the Augment description deck was written. The version of MP published in TOMS and formerly available from the ACM Algorithms Distribution service (version 770217) did not include Augment interface routines, but a version of MP which includes them (version 780802) is now available from the Algorithms Distribution service. The most recent version of MP may be obtained from the author at the address given on the title page of this User's Guide.

In 1979 the storage allocation scheme was improved, rounding options were implemented for many MP routines, dependence on Fortran library routines such as EXP and ALOG and on REAL arithmetic was eliminated, many routines were modified to allow packed as well as unpacked multiple-precision arguments, and corresponding changes were made in the Augment description deck. For more details, see Section 8.1. The MP User's Guide (Second Edition) was extensively revised to produce the third edition, and at the same time it was converted from upper-case to duo-case. Only minor changes and corrections have been made since then.

For an introduction to the design and philosophy of MP, see - R. P. Brent, 'A Fortran multiple-precision arithmetic package', ACM Trans. Math. Software 4 (March 1978), 57-70. Additional details are given in 'Algorithm 524 - MP, a Fortran multiple-precision arithmetic package', ibid, 71-81, and in Section 6 below. The mathematical theory behind the algorithms used in the MP package is described in - R. P. Brent, 'Unrestricted algorithms for elementary and special functions', Information Processing 80, North Holland, 1980, 613-619.

As mentioned above, a preprocessor (Augment) which facilitates the use of the MP package is available. See - R. P. Brent, J. A. Hooper and J. M. Yohe, 'An Augment interface for Brent's multiple-precision arithmetic package', ACM Trans. Math. Software 6 (1980), 146-149, and Sections 4 and 5 below.

Motivation for certain details of our implementation of T-digit, base B arithmetic is given in - C. H. Reinsch, 'Principles and preferences for computer arithmetic', ACM Signum Newsletter 14, 1 (March 1979), 12-27. In designing the Augment interface we have been influenced by the IFIP Working Group 2.5 proposals - see B. Ford (ed.), 'Parameterization of the environment for transportable numerical software', ACM Trans. Math. Software 4 (1978), 100-103. See also - W. S. Brown and S. I. Feldman, 'Environment parameters and basic functions for floating-point computations', ACM Signum Newsletter 14, 1 (March 1979), 42-45.

In the future we hope to implement rounding options for more MP routines, and write a multiple-precision interval arithmetic package which uses MP and takes advantage of the directed rounding options. An Augment interface to facilitate the use of the interval arithmetic package would be provided. (A similar project has been attempted by John P. Jeter - see his 'A variable precision interval data type extension to Fortran', M. Sc. thesis, Univ. of South West Louisiana, July 1979.)

A never-ending project is to implement multiple-precision versions of ever more special functions, and to improve the efficiency of those multiple-precision routines already implemented.

Correspondence concerning MP should be addressed to Prof. R. P. Brent at the address given on the title page. Reports of bugs, information on applications for which the MP package has proved useful, and suggestions for the future development of the package are particularly welcome.

## 1.5 Summary of useful MP routines

We mention here the names of those MP routines which are likely to be of interest to someone using the MP package without the aid of the Augment interface. Routines which are called by other MP routines but are unlikely to be called directly by a user of the package are omitted. For the user of Augment, the summary given in Section 5.2 will be more useful. Details of all MP routines may be found in Section 6.

Basic arithmetic

> MPADD, MPADDI, MPADDQ, MPDIV, MPDIVI, MPMUL, MPMULI, MPMULQ, MPREC, MPSUB

Powers and roots

> MPPWR, MPPWR2, MPQPWR, MPROOT, MPSQRT

Elementary functions

> MPASIN, MPATAN, MPATN2, MPCIS, MPCOS, MPCOSH, MPEXP, MPEXP1, MPLG10, MPLN, MPLNI, MPLNS, MPSIN, MPSINH, MPTAN, MPTANH

Special functions

> MPBESJ, MPDAW, MPEI, MPERF, MPERFC, MPGAM, MPGAMQ, MPLI, MPLNGM

Constants

> MPBERN, MPEPS, MPEUL, MPMAXR, MPMINR, MPPI, MPZETA

Input and output

> MPFIN, MPFOUT, MPIN, MPOUT, MPUNFR, MPUNFW

Conversion

> MPCAM, MPCDM, MPCIM, MPCMD, MPCMDE, MPCMI, MPCMR, MPCMRE, MPCQM, MPCRM

Comparison

> MPCMPA, MPCMPD, MPCMPI, MPCMPQ, MPCMPR, MPCOMP, MPEQ, MPGE, MPGT, MPLE, MPLT, MPNE

General utility routines

> MPABS, MPCEIL, MPCHEB, MPCHEV, MPCMF, MPCMIM, MPDIGS, MPDIM, MPFLOR, MPGCDA, MPGCDB, MPINIT, MPMAX, MPMIN, MPMOD, MPNEG, MPPACK, MPPARA, MPPARB, MPPOLY, MPSETR, MPSET2, MPSIGN, MPSTR, MPUNPK

Example and test programs

> EXAMPLE, JACOBI, TEST, TEST2

2. EXAMPLE PROGRAM

The  following  program  illustrates  the  straightforward use of the MP
package without the aid of Augment.  An example using Augment  is  given
in Section 4.3.

```
      C $$                 ******  EXAMPLE  ******
      C
      C THIS PROGRAM COMPUTES PI AND EXP(PI*SQRT(163/9)) TO 100
      C DECIMAL PLACES, AND EXP(PI*SQRT(163)) TO 90 DECIMAL PLACES,
      C AND WRITES THEM ON LOGICAL UNIT 6.
      C
      C CORRECT OUTPUT (EXCLUDING HEADINGS) IS AS FOLLOWS
      C
      C                   3.1415926535897932384626433832795028841971693993751O
      C                    5820974944592307816406286208998628034825342117068O
      C             640320.00000000060486373504901603947174181881853947577148
      C                    5760366591819465221825828694253634081582264647759O
      C 262537412640768743.9999999999925007259719818568887935385633733699086
      C                    2707537410378210647910118607312951181346
      C
      C WORKING SPACE IN BLANK COMMON, PARAMETERS IN
      C COMMON /MPCOM/ (WHICH HAS LENGTH 23)
            COMMON R
            COMMON /MPCOM/ B, T, M, LUN, MXR, SPTR, MXSPTR, DUMMY
            INTEGER B, T, M, LUN, MXR, SPTR, MXSPTR, DUMMY(16), R(500)
      C
      C WE HAVE T .LE. 62 IF WORDLENGTH AT LEAST 16 BITS, AND WORKING SPACE
      C IS AT MOST 500 WORDS (LESS IF WORDLENGTH IS GREATER THAN 16 BITS).
      C
      C VARIABLES NEED T+2 .LE. 64 WORDS AND ALLOW 110 CHARACTERS FOR
      C DECIMAL OUTPUT
            INTEGER PI(64), X(64), C(110)
      C
      C CALL MPSET2 TO SET OUTPUT LOGICAL UNIT = 6 AND EQUIVALENT
      C NUMBER OF DECIMAL PLACES TO AT LEAST 110.  THE LAST THREE
      C PARAMETERS ARE THE DIMENSIONS OF PI (OR X) AND THE LOWER
      C AND UPPER INDICES OF BLANK COMMON AVAILABLE TO MP.
            CALL MPSET2 (6, 110, 64, 1, 500)
      C
      C COMPUTE MULTIPLE-PRECISION PI
            CALL MPPI(PI)
      C
      C CONVERT TO PRINTABLE FORMAT (F110.100) AND WRITE
            CALL MPOUT (PI, C, 110, 100)
            WRITE (LUN, 10) B, T, C
         10 FORMAT (32H1EXAMPLE OF MP PACKAGE,   BASE =, I9,
          $  12H,   DIGITS =, I4 /// 11H PI TO 100D //
          $  11X, 60A1 / 21X, 50A1)
      C
      C SET X = SQRT(163/9), THEN MULTIPLY BY PI
            CALL MPQPWR (163, 9, 1, 2, X)
            CALL MPMUL (X, PI, X)
```

```
C SET X = EXP(X)
      CALL MPEXP (X, X)
C
C CONVERT TO PRINTABLE FORMAT AND WRITE
      CALL MPOUT (X, C, 110, 100)
      WRITE (LUN, 20) C
   20 FORMAT (/ 28H EXP(PI*SQRT(163/9)) TO 100D //
     $         11X, 60A1 / 21X, 50A1)
C
C SET X = X**3 = EXP(PI*SQRT(163))
      CALL MPPWR (X, 3, X)
C
C WRITE IN FORMAT F110.90
      CALL MPOUT (X, C, 110, 90)
      WRITE (LUN, 30) C
     $         1X, 70A1 / 21X, 40A1)
      WRITE (LUN, 40) MXSPTR
   40 FORMAT (/ 21H END OF EXAMPLE, USED, I4,
     $         23H WORDS OF WORKING SPACE //)
      STOP
      END
```

# 3. PARAMETERS IN COMMON /MPCOM/

## 3.1 Description of the parameters

The labelled COMMON block, COMMON /MPCOM/, contains the following 23 parameters which are used by various routines in the MP package and referred to in Section 6. In the description below, 'default value' means the value set by a call to MPSET2 (or MPINIT or MPSET, since they call MPSET2). All the parameters are of type INTEGER.

(1) BASE or B     The base of the multiple-precision number representation. B must be at least 2, and $8*B**2-1$ must be representable as a single-precision integer. The default value is the largest power of two satisfying this condition.

(2) NUMDIG or T   The number of multiple-precision digits (must be at least 2). The default value is sufficient to give accuracy equivalent to at least DECPL decimal places.

(3) MAXEXP or M   The maximum exponent of multiple-precision numbers. Exponents in the range 1-M, ... , +M are allowed, so numbers from $B**(-M)$ to almost $B**M$ are representable. M must be greater than T. The default setting and maximum allowable value is int (MXINT/4).

(4) LUN           The logical unit for output by routines MPFOUT, MPOUTF, MPDUMP, MPERR, MPERRM etc. The default value is the first argument of the last call to MPSET2 (or 6 if MPINIT as distributed with the MP package is called).

(5) MXR           Words MNSPTR to MXR of blank COMMON are used as working space by the MP package. MXR-MNSPTR must be sufficiently large (see Section 6). The default values of MNSPTR and MXR are the 4-th and 5-th arguments of the last call to MPSET2 (or 1 and 500 if MPINIT is used as distributed with the MP package).

(6) SPTR          The current pointer to the top of the working storage stack (plus one). SPTR should be in the range MNSPTR <= SPTR <= MXR + 1. The default initial setting is MNSPTR.

(7) MXSPTR        The maximum value of SPTR during the current run. The default initial setting is MNSPTR.

(8) MNSPTR        The minimum value of SPTR during the current run. See the description of MXR above.

(9) MXEXPN        The maximum exponent of any multiple-precision number during the current run. The default initial setting is -M. Set to M+1 if multiple-precision overflow occurs.

(10) MNEXPN       The minimum exponent of any multiple-precision number
                  during the current run. The default initial setting is
                  M+1. Set to -M if a multiple-precision underflow
                  occurs.

(11) RNDRL        Integer in the range 0, ... , 3 indicating the type of
                  rounding to be performed. Generally,

                     0 means truncated (chopped) arithmetic,
                     1 means rounded arithmetic,
                     2 means round down (towards -infinity),
                     3 means round up (towards +infinity).

                  However, these interpretations do not always apply.
                  For details of the effect of RNDRL, see Section 6. The
                  default setting is 0 (truncated arithmetic).

(12) KTUNFL       Counter for the number of multiple-precision underflows
                  which have occurred. The default initial setting is
                  zero.

(13) MXUNFL       If positive, execution is terminated after MXUNFL
                  multiple-precision underflows have occurred. If zero,
                  any number of multiple-precision underflows are allowed
                  (this is the default).

(14) DECPL        A conservative estimate of the 'equivalent' number of
                  floating-point decimal places. The default setting is
                  the second argument in the last call to MPSET2 (or 43
                  if MPINIT is used as distributed). The default
                  settings of B and T satisfy

                     B**(T - 1) >= 10**(DECPL - 1)

(15) MT2          The dimension of arrays used for multiple-precision
                  variables. The default setting is the third argument
                  in the last call to MPSET2 (or 27 if MPINIT is used as
                  distributed).

(16) MXINT        The largest single-precision integer of the form
                  2**k - 1 such that INTEGER arithmetic with operands and
                  results in the range -MXINT, ... , +MXINT is performed
                  exactly. Usually k = (wordlength in bits) - 1. See
                  also the description of MPLARG in Section 6.

(17) EXWID        The number of characters in the exponent field for
                  output of multiple-precision numbers using MPFOUT. The
                  default value is 6. (Since one position is taken by the
                  exponent character EXPCH and one by the sign, this
                  means that exponents in the range -9999 to +9999 are
                  allowable with the default setting.)

(18) INRECL       The input record length assumed by MPFIN. Must not
                  exceed 80 (the default value) unless MPFIN and MPCHK
                  are modified.

(19) INBASE      The base of 'decimal' numbers read by MPIN, MPFIN  etc.
                 (e.g.  INBASE = 8 means octal input, INBASE = 16 means
                 hexadecimal input).  Must be in the range 2, ... , 16.
                 The default setting is ten, for decimal input.

(20) OUTBAS      The base for 'decimal' output by  MPOUT,  MPFOUT  etc.
                 Default  setting ten, must be in the range 2, ... , 16.

(21) EXPCH       The exponent character used by MPFOUT.  Must not  be  a
                 digit, sign or blank.  The default setting is 'E'.

(22) CHWORD      The number of characters per  single-precision  integer
                 word.  The default setting is determined by MPUPW.

(23) ONESCP      1 or 0 if the machine  uses  one's  complement  integer
                 arithmetic,  0  otherwise.  The default setting is
                 determined by MPSET2 and MPUPW so that MPUPW works.

3.2 Methods of setting the parameters

It would, of course, be possible to declare the parameters B, T, M, LUN,
... ,  ONESCP  in  COMMON  /MPCOM/  and  initialize  them by assignment
statements or by a BLOCK DATA subprogram.  However, this method  is  not
recommended.   Instead, we recommend calling MPINIT or MPSET2 to set the
default values (see above and Section 6), and then using MPPARB or PARAM
to change the setting of any parameters for which the default values are
not desired.

For  example,  to  use  the  MP  package  with 20 place rounded floating
decimal arithmetic, we could do the following -

        Without using Augment                Using Augment

        CALL MPINIT (MP)                     INITIALIZE MP
        CALL MPPARB (20, 'NUMDIG')           PARAM ('NUMDIG') = 20
        CALL MPPARB (10, 'BASE')             PARAM ('BASE') = 10
        CALL MPPARB (1, 'RNDRL')             PARAM ('RNDRL') = 1

MPPARC may be used instead of MPPARB.   It is less mnemonic, but slightly
faster, so might be preferred inside a loop. For details see Section  6.

The   current   setting   of   the parameters may be determined by MPPARA or
PARAM.  For example, MPPARA ('NUMDIG') will return the current number of
multiple-precision  digits.   For  users of Augment, PARAM ('NUMDIG') on
the right-hand side of an assignment  statement  will  return  the  same
information.

Augment users may prefer to set the first  three  parameters  in  COMMON
/MPCOM/  with  the  field  functions  BASE,  NUMDIG and MAXEXP instead of
PARAM.  For example, if X is a variable of type MULTIPLE, the number  of
multiple-precision digits may be set to 20 by the statement

        NUMDIG (X) = 20

For further details, see Section 5.2.

3.3 Efficiency considerations

For efficiency choose B fairly large, subject to the restrictions given in Section 1.2.  For example, if the wordlength is

        48 bits, could use B = 4194304 or 1000000,
        36 bits, could use B = 65536 or 100000,
        32 bits, could use B = 16384 or 10000,
        24 bits, could use B = 1024 or 1000,
        18 bits, could use B = 128 or 100,
        16 bits, could use B = 64 or 10,
        12 bits, could use B = 16 or 10.

Of course, for special purposes such as simulation of decimal or binary arithmetic, B may be set as small as 2.  If a significant amount of decimal input/output is to be performed, it may be best to choose B a power of ten (MPIN and MPOUT will take advantage of this).

Avoid multiplication and division by multiple-precision numbers, as these take O(T**2) operations, whereas multiplication and division by (single-precision) integers take O(T) operations.  See the descriptions of MPDIV, MPDIVI, MPMUL and MPMULI in Section 6.

The default setting of RNDRL is 0 (i.e. truncated arithmetic). The use of other values of RNDRL is more expensive in both time and space.

# 4. THE AUGMENT INTERFACE

## 4.1 Introduction

Augment is a preprocessor which allows the introduction of nonstandard types (e.g. multiple-precision variables) into Fortran programs. For details, see - F.D. Crary, 'A versatile precompiler for nonstandard arithmetics', ACM Trans. Math. Software 5 (June 1979), 204-217, and the references given there.

A 'description deck' has been written to enable Augment to be used in conjunction with the MP package. This greatly simplifies the task of writing a program for a multiple-precision computation, or converting a single (or double) precision routine to multiple precision.

For example, if Augment is used we can declare X, Y and Z to be multiple-precision variables by a declaration such as

```
      MULTIPLE X, Y, Z
```
or
```
      IMPLICIT MULTIPLE (A-H, O-Z)
```

and subsequently write expressions such as

```
      X = Y + Z*EXP(X+1)/Y
```

instead of the equivalent

```
      CALL MPADDI (X, 1, TEMP)
      CALL MPEXP (TEMP, TEMP)
      CALL MPMUL (Z, TEMP, TEMP)
      CALL MPDIV (TEMP, Y, TEMP)
      CALL MPADD (Y, TEMP, X)
```

The Augment interface can be used with MP version 780420, or later versions. For more details, see - R. P. Brent, J. A. Hooper and J. M. Yohe, 'An Augment interface for Brent's multiple-precision arithmetic package', ACM Trans. Math. Software 6 (1980), 146-149.

A convenient summary of the operations implemented in the MP package and the methods of invoking them via Augment is given in Section 5.2.

4.2 The description deck

The 'description deck' which describes the MP package is as follows.

```
COMMENT   AUGMENT DESCRIPTION DECK FOR THE MULTIPLE-PRECISION ARITHMETIC
          PACKAGE OF R. P. BRENT.

          THREE TYPES OF VARIABLE ARE DEFINED HERE -

               MULTIPLE   (STANDARD MULTIPLE-PRECISION NUMBERS),
               MULTIPAK   (PACKED MULTIPLE-PRECISION NUMBERS), AND
               INITIALIZE (USED ONLY AS A DEVICE TO PERSUADE
                          AUGMENT TO INITIALIZE THE MP PACKAGE).

          WORKING SPACE SHOULD BE ALLOCATED AND THE MP PACKAGE
          INITIALIZED BY THE DECLARATION

                    INITIALIZE MP

          IN THE MAIN PROGRAM.

          THIS DESCRIPTION DECK ASSUMES THAT EACH MP NUMBER REQUIRES NO
          MORE THAN 27 WORDS (14 IN PACKED FORMAT).  THIS IS SUFFICIENT
          FOR ABOUT 43 DECIMAL PLACE ACCURACY ON A 16-BIT MACHINE, AND
          HIGHER ACCURACY ON A MACHINE WITH A LONGER WORDLENGTH.

          SEE COMMENTS IN ROUTINE MPINIT FOR THE METHOD OF CHANGING THE
          PRECISION, ALSO REGARDING COMMON DECLARATIONS.

*DESCRIBE MULTIPAK

DECLARE INTEGER (14),  KIND SAFE SUBROUTINE, PREFIX MPK

CONVERSION CTP (CIM, INTEGER, $, UPWARD),  CTP (CAM, HOLLERITH)

SERVICE COPY (=MPSTR)

*DESCRIBE MULTIPLE

DECLARE INTEGER (27),  KIND SAFE SUBROUTINE, PREFIX MP

OPERATOR + (, NULL UNARY, PRV, $),  - (NEG, UNARY),
         + (ADD, BINARY3, PRV, $, $, $, COMM),  * (MUL),
         - (SUB,,,,,, NONCOMM),  / (DIV),  ** (PWR2),

         + (ADDI,,,, INTEGER),  * (MULI),  / (DIVI),  ** (PWR),
         * (IMUL,,, INTEGER, $, $, NONCOMM),

         .EQ. (EQ, BINARY2, PRV, $, LOGICAL, COMM),  .NE. (NE),
         .GE. (GE,,,,, NONCOMM),  .GT. (GT),  .LE. (LE),  .LT. (LT),

         + (, NULL UNARY, PRV, MULTIPAK),
         - (NEG, UNARY, PRV, MULTIPAK),
         + (KADD, BINARY3, PRV, MULTIPAK, MULTIPAK, $, COMM),
         * (KMUL),  - (KSUB,,,,,, NONCOMM),  / (KDIV),  ** (PWR2),
```

```
                * (KMLI,,,, INTEGER),   / (KDVI),   ** (PWR),
                * (KIML,,, INTEGER, MULTIPAK, $, NONCOMM),

                .EQ. (EQ, BINARY2, PRV, MULTIPAK, LOGICAL, COMM), .NE. (NE),
                .GE. (GE,,,,, NONCOMM), .GT. (GT), .LE. (LE), .LT. (LT)

TEST     MPSIGA (SIGA, INTEGER)

FIELD    SGN (SIGA, SIGB, ($), INTEGER),  EXPON (EXPA, EXPB),
         BASE (BASA, BASB), NUMDIG (DIGA, DIGB), MAXEXP (MEXA, MEXB),
         DIGIT (DGA, DGB, ($, INTEGER)),
         PARAM (PARA, PARB, (HOLLERITH)), FIO (FIN, FOUT, (INTEGER),
         $), UNFIO (UNFR, UNFW),
         EXPON (EXPA, EXPB, (MULTIPAK), INTEGER), NUMDIG (DIGA, DIGB),
         SGN (SIGA, SIGB)

FUNCTION ABS (ABS, ($), $), DABS (ABS), ASIN (ASIN), DASIN (ASIN),
         ARSIN (ASIN), DARSIN (ASIN), ARCSIN (ASIN), ATAN (ATAN),
         DATAN (ATAN), ARTAN (ATAN), DARTAN (ATAN), ARCTAN (ATAN),
         CMF (CMF), CEIL (CEIL), FLOOR (FLOR), CMIM (CMIM),
         COS (COS), DCOS (COS), COSH (COSH), DCOSH (COSH),
         DAW (DAW), DDAW (DAW), EI (EI), DEI (EI), ERF (ERF),
         DERF (ERF), ERFC (ERFC), DERFC (ERFC), EXP (EXP),
         DEXP (EXP), EXP1 (EXP1), FRAC (CMF), GAM (GAM),
         DGAM (GAM), GAMMA (GAM), DGAMMA (GAM), AINT (CMIM),
         DINT (CMIM), LI (LI), DLI (LI), LN (LN), LOG (LN),
         ALOG (LN), DLOG (LN), LOG10 (LG10), ALOG10 (LG10),
         DLOG10 (LG10), LNGM (LNGM), ALNGM (LNGM), DLNGM (LNGM),
         LNGS (LNGS), LNS (LNS), REC (REC), SIN (SIN), DSIN (SIN),
         SINH (SINH), DSINH (SINH), SQRT (SQRT), DSQRT (SQRT),
         TAN (TAN), DTAN (TAN), TANH (TANH), DTANH (TANH),

         SNGL (STR), DBLE (STR),

         ART1 (ART1, (INTEGER)), LN (LNI), LNI (LNI), LOG (LNI),
         ALOG (LNI), DLOG (LNI), ZETA (ZETA),

         CAM (CAM), CAM (CAM, (HOLLERITH)),

         MAX (MAX, ($, $)), AMAX1 (MAX), DMAX1 (MAX), MIN (MIN),
         AMIN1 (MIN), DMIN1 (MIN), DIM (DIM), DDIM (DIM),
         GCD (GCDA), ATAN2 (ATN2), DATAN2 (ATN2), ARTAN2 (ATN2),
         MOD (MOD), AMOD (MOD), DMOD (MOD),
         SIGN (SIGN), DSIGN (SIGN),

         BESJ (BESJ, ($, INTEGER)), ROOT (ROOT),

         MPINF (INF (SUBROUTINE), ($, INTEGER, INTEGER, HOLLERITH),
         LOGICAL), MPOUTF (OUTF (SUBROUTINE)),
         MPINF (INF (SUBROUTINE), ($, INTEGER, INTEGER, INTEGER)),
         MPOUTF (OUTF (SUBROUTINE)),

         INTEGR (INTG, ($), LOGICAL),
```

```
        COMP (COMP, ($, $),   INTEGER),   CMPA (CMPA),
        COMP (CMPI, ($, INTEGER)),  COMP (CMPR, ($, REAL)),
        COMP (CMPD, ($, DOUBLE PRECISION)),
        COMP (CMPQ, ($, INTEGER, INTEGER)),

        ADDQ (ADDQ, ($, INTEGER, INTEGER), $),  MULQ (MULQ),
        QPWR (QPWR, (INTEGER, INTEGER, INTEGER, INTEGER)),
        CQM (CQM, (INTEGER, INTEGER)),  CTM (CQM),

        GAM (GAMQ),   DGAM (GAMQ),   GAMQ (GAMQ),

        BERN (BERN, (INTEGER, INTEGER), MULTIPAK),

        INT (CMI (SUBROUTINE),   ($),  INTEGER),
        IDINT (CMI (SUBROUTINE)),  IFIX (CMI (SUBROUTINE)),

        ABS (ABS, (MULTIPAK), MULTIPAK), ATAN (ATAN,, $), COS (COS),
        COSH (COSH),  EXP (EXP),  LN (LN),  LOG (LN),  ALOG (LN),
        LOG10 (LG10),  ALOG10 (LG10),  SIN (SIN),  SINH (SINH),
        SQRT (SQRT),  TAN (TAN),  TANH (TANH),
        EXP1 (EXP1), LNS (LNS),

        MAX (MAX,  (MULTIPAK, MULTIPAK)),  MIN (MIN),  DIM (DIM),
        ATAN2 (ATN2),  MOD (MOD),  SIGN (SIGN),
        ROOT (ROOT, (MULTIPAK, INTEGER))

CONVERSION CTM (CDM, DOUBLE PRECISION, $, UPWARD),  CTM (CIM, INTEGER),
        CTM (CRM, REAL),  CTM (UNPK, MULTIPAK),  CTM (CAM, HOLLERITH),
        CTD (CMD (SUBROUTINE),  $, DOUBLE PRECISION, DOWNWARD),
        CTI (CMI (SUBROUTINE),, INTEGER),  CTR (CMR (SUBROUTINE),,
        REAL),  CTP (PACK,, MULTIPAK)

SERVICE COPY (STR)

*DESCRIBE INITIALIZE

DECLARE INTEGER (1),  KIND SAFE SUBROUTINE, PREFIX MPI
SERVICE COPY (STR),   INITIAL (NIT)

COMMENT    END OF AUGMENT DESCRIPTION DECK FOR MP PACKAGE
```

4.3 JACOBI - an example of the use of Augment

The program which follows illustrates the use of the MP package with the Augment interface. It is quite straightforward, and does not by any means illustrate all the possibilities provided by the Augment interface.

```
< Machine-dependent statement(s) to execute Augment >
PRINT SUPPRESS
< Machine-dependent statement(s) to insert the description deck >

*BEGIN
*DISABLE PRINT, OUTPUT
*ENABLE SOURCE

< Machine-dependent statement(s) to (later) execute Fortran compiler >

C
C PROGRAM TO VERIFY AN IDENTITY OF JACOBI USING THE MP
C PACKAGE AND AUGMENT.
C
C THE PROGRAM READS A NUMBER X IN FREE-FIELD FORMAT ACCEPTABLE TO
C MPIN.  IF X IS NON-POSITIVE IT HALTS.  OTHERWISE IT COMPUTES
C AND PRINTS FN(X), FN(1/X) AND (FN(X)-FN(1/X))/FN(X),
C WHERE  FN(X) IS THE SUM FROM N = -INFINITY TO +INFINITY OF
C SQRT(X)*EXP(-PI*(N*X)**2).
C THE IDENTITY VERIFIED IS
C                        FN(X) = FN(1/X)
C
C DECLARE VARIABLES AND INITIALIZE MP PACKAGE.  ON SOME SYSTEMS BLANK
C COMMON MUST BE DECLARED HERE - SEE COMMENTS IN ROUTINE MPINIT.
C
      MULTIPLE X, F1, F2, FN
      INITIALIZE MP
C
C SET INPUT RECORD LENGTH TO 72 (TO AVOID READING
C SEQUENCE NUMBERS), DEFAULT VALUE IS 80.
C
      PARAM (6HINRECL) = 72
C
C READ MP X FROM UNIT 5 IN FREE FORMAT, STOP IF X NOT POSITIVE.
C
   10 X = FIO (5)
      IF (X.LE.0) STOP
C
C WRITE HEADING, X, FN(X), AND FN(1/X) TO 40S IN STANDARD FORMAT
C
      WRITE (6, 20)
   20 FORMAT (//41H X, FN(X), FN(1/X), (FN(X)-FN(1/X))/FN(X)//)
      FIO (40) = X
      F1 = FN(X)
      FIO (40) = F1
      F2 = FN(1/X)
      FIO (40) = F2
```

```
C WRITE (F1-F2)/F1 TO 6 SIGNIFICANT FIGURES.
C NOTE THAT AN MP EXPRESSION CAN BE WRITTEN.
C
      FIO (6) = (F1 - F2)/F1
      GO TO 10
      END
C
< Machine-dependent statement(s) to (later) execute Fortran compiler >
C
      FUNCTION FN(X)
C
C RETURNS FN(X) = THE SUM FROM N = -INFINITY TO +INFINITY OF
C SQRT(X)*EXP(-PI*(N*X)**2), ASSUMING X POSITIVE.
C USES THE OBVIOUS METHOD, SO SLOW IF X SMALL.
C NOTE THAT X AND FN ARE BOTH TYPE MULTIPLE.
C TO ILLUSTRATE MIXED-MODE ARITHMETIC, SOME LOCAL
C VARIABLES ARE DECLARED TO BE PACKED.
C
      MULTIPLE FN, X
      MULTIPAK TM, FAC, PR
      IF (X.LE.0) CALL MPERR
      FN = 0
C
C AUGMENT CAN DEAL WITH THE FOLLOWING EXPRESSION AS IT KNOWS THAT X
C IS TYPE MULTIPLE, SO CALLS MPCAM TO CONVERT 'PI' TO MULTIPLE.
C
      TM = EXP(-2HPI*X*X)
      PR = TM
      FAC = TM**2
C
C LOOP TO SUM INFINITE SERIES
C WARNING - NUMBER OF ITERATIONS IS PROPORTIONAL TO 1/X
C
   10 FN = FN + TM
      PR = PR*FAC
      TM = TM*PR
C
C TEST FOR CONVERGENCE BY COMPARING EXPONENTS OF FN AND TM.
C WE COULD ALSO HAVE SAVED THE OLD VALUE OF FN AND SEEN IF
C STATEMENT 10 CHANGED IT.
C
      IF (EXPON(FN)-EXPON(TM).LT.NUMDIG(X)) GO TO 10
      FN = SQRT(X)*(2*FN+1)
      RETURN
      END
*END

<Machine-dependent statements to add Augment output file to
 the runstream, link compiled Fortran routines with precompiled
 library of MP routines, and execute with the following data >
.5
.3
1.+1
1.234567890123456789012345678901234567890123456789
0
```

# 5. SUMMARY FOR AUGMENT USERS

## 5.1 Notes

We summarize as briefly as possible how MP routines may be invoked using the Augment interface. Under 'Invocation' we give the typical way of invoking the operation described under 'Description'. For further details of calling sequences etc., the column 'Ref' gives a reference to the relevant MP routine, and descriptions of these routines may be found in Section 6. The column headed 'Res' gives the type of the result of each operation.

Types are indicated by one-letter abbreviations,

      D = double-precision real (DOUBLE PRECISION),
      G = packed or unpacked multiple (MULTIPAK or MULTIPLE)
          (indicated by 'MP*' in the column headed 'Description'),
      H = packed Hollerith (as in 'ABCDEF'),
      I = single-precision integer (INTEGER),
      L = Boolean (LOGICAL),
      M = unpacked multiple (MULTIPLE),
      P = packed multiple (MULTIPAK),
      R = single-precision real (REAL).

Arguments are written as XA, XB, XC, ... , where X indicates the type, as above. XDUMMY means a dummy argument of the type indicated by X. Such arguments serve no useful purpose, but Augment expects them.

The conversion routines may be invoked implicitly by assignment statements, for example -

      MR = IA

will be translated by Augment into a call to MPCIM.

Packed (i.e. type MULTIPAK) and unpacked (i.e. type MULTIPLE) multiple-precision variables may be mixed in simple arithmetic expressions. For example, in the statement -

      A = B + C/(D - E*F)

any combination of packed and unpacked multiple-precision variables is permissible. Augment will generate calls on the appropriate conversion routines. We do not recommend using packed multiple-precision variables in expressions involving function calls, without explicit type conversion (using CTM and/or CTP), or unless it is mentioned as permissible in the summary below. Augment will not always generate calls on the appropriate conversion routines.

Field functions may occur on either side of an assignment statement. For example -

      PARAM ('MXUNFL') = 10

would set the maximum allowable number of MP underflows to 10.

## 5.2 Summary

```
        Description                      Res Invocation          Ref

  Addition

    Sum of two MP numbers                M   MA + MB             MPADD
    Sum of two packed MP numbers         M   PA + PB             MPKADD
    Sum of MP number and an integer      M   MA + IB             MPADDI
    Sum of MP number and rational IB/IC  M   ADDQ (MA, IB, IC)   MPADDQ

  Division

    Quotient of two MP numbers           M   MA / MB             MPDIV
    Quotient of two packed MP numbers    M   PA / PB             MPKDIV
    Quotient of MP number and an integer M   MA / IB             MPDIVI
    Quotient of packed MP and integer    M   PA / IB             MPKDVI

  Multiplication

    Product of two MP numbers            M   MA * MB             MPMUL
    Product of two packed MP numbers     M   PA * PB             MPKMUL
    Product of MP number and integer     M   MA * IB             MPMULI
    Product of packed MP and integer     M   PA * IB             MPKMLI
    Product of integer and MP number     M   IA * MB             MPIMUL
    Product of integer and packed MP     M   IA * PB             MPKIML
    Product of MP and rational IB/IC     M   MULQ (MA, IB, IC)   MPMULQ

  Reciprocal

    Reciprocal of MP number              M   REC (MA)            MPREC

  Subtraction

    Difference of two MP numbers         M   MA - MB             MPSUB
    Difference of two packed MP numbers  M   PA - PB             MPKSUB

  Powers and roots

    Raise MP* number to integer power    M   GA ** IB            MPPWR
    Raise MP number to MP power          M   MA ** MB            MPPWR2
    Raise packed MP to packed MP power   M   PA ** PB            MPPWR2
    Raise rat. IA/IB to rat. power IC/ID M   QPWR(IA,IB,IC,ID)   MPQPWR
    IB-th root of MP* number             M   ROOT (GA, IB)       MPROOT

  Elementary functions

    Arcsin of MP number (in radians)     M   ASIN (MA)           MPASIN
    Arctan of MP* number (in radians)    M   ATAN (GA)           MPATAN
    Arctan of 1/IA, IA > 1               M   ART1 (IA)           MPART1
    Arctan of MA/MB (both MP numbers)    M   ATAN2 (MA, MB)      MPATN2
    Arctan of PA/PB (both packed MP)     M   ATAN2 (PA, PB)      MPATN2
    Cosine of MP* number (radians)       M   COS (GA)            MPCOS
    Hyperbolic cosine of MP* number      M   COSH (GA)           MPCOSH
```

```
            Description                     Res Invocation          Ref
```

Elementary functions continued

```
    Exponential of MP* number              M   EXP (GA)             MPEXP
    (Exponential - 1) of MP* number        M   EXP1 (GA)            MPEXP1
    Natural logarithm of MP* number        M   LOG (GA)             MPLN
    Logarithm (base 10) of MP* number      M   LOG10 (GA)           MPLG10
    Natural logarithm of (1 + MP* number)  M   LNS (GA)             MPLNS
    Natural logarithm of integer > 0       M   LOG (IA)             MPLNI
    Sine of MP* number (radians)           M   SIN (GA)             MPSIN
    Hyperbolic sine of MP* number          M   SINH (GA)            MPSINH
    Square root of MP* number              M   SQRT (GA)            MPSQRT
    Tangent of MP* number (radians)        M   TAN (GA)             MPTAN
    Hyperbolic tangent of MP* number       M   TANH (GA)            MPTANH
```

Special functions

```
    Bessel function of first kind,
       MP argument MA, integer order IB    M   BESJ (MA, IB)        MPBESJ
    Dawson's integral of MP argument       M   DAW (MA)             MPDAW
    Exponential integral of MP argument    M   EI (MA)              MPEI
    Error function of MP number            M   ERF (MA)             MPERF
    Complementary error function           M   ERFC (MA)            MPERFC
    Gamma function of MP number            M   GAM (MA)             MPGAM
    Gamma function of rational IA/IB       M   GAM (IA, IB)         MPGAMQ
    GCD of MP numbers which are integers   M   GCD (MA, MB)         MPGCDA
    Logarithmic integral of MP number      M   LI (MA)              MPLI
    Logarithm of Gamma function            M   LNGM (MA)            MPLNGM
    Riemann zeta function of integer > 1   M   ZETA (IA)            MPZETA
```

Constants

```
    Bernoulli numbers (result is array)    P   BERN (IA, IB)        MPBERN
    MP machine precision                   M   CTM ('EPS')          MPEPS
    Euler's constant                       M   CTM ('EUL')          MPEUL
    Largest positive MP number             M   CTM ('MAXR')         MPMAXR
    Smallest positive MP number            M   CTM ('MINR')         MPMINR
    Pi                                     M   CTM ('PI')           MPPI
    Decimal constants, e.g. 6.3E5          M   CTM ('6.3E5$')       MPCAM
```

Input and output

```
    Read MP number from unit IA in
       free format                         M   FIO (IA)             MPFIN

    Write MP number MB with IA decimal
       places on default output unit           FIO (IA) = MB        MPFOUT

    Read IB words from unit IC under
       format HD and convert to MP         L   MPINF(MA,IB,IC,HD)
       number MA, result error flag                                 MPINF
```

```
            Description                    Res Invocation           Ref

  Input and output continued

      Write IB words representing MP
        number MA on default unit, with
        IC places after decimal point,     L   MPOUTF(MA,IB,IC,HD)
        under format HD, result error flag                          MPOUTF

      Read MP number from unit IA,
        unformatted                        M   UNFIO (IA)            MPUNFR

      Write MP number MB on unit IA,
        unformatted                            UNFIO (IA) = MB       MPUNFW

  Conversion

      Double-precision to multiple         M   CTM (DA)              MPCDM
      Integer to multiple                  M   CTM (IA)              MPCIM
      Real to multiple                     M   CTM (RA)              MPCRM
      Rational IA/IB to multiple           M   CTM (IA, IB)          MPCQM
      Packed multiple to unpacked multiple M   CTM (PA)              MPUNPK
      Packed Hollerith to multiple         M   CTM (HA)              MPCAM
      Packed Hollerith stored in integer
        or integer array to multiple       M   CAM (IA)              MPCAM
      Multiple to double-precision         D   CTD (MA)              MPCMD
      Multiple to integer (truncated)      I   CTI (MA)              MPCMI
      Multiple to real                     R   CTR (MA)              MPCMR
      Unpacked multiple to packed multiple P   CTP (MA)              MPPACK

  Comparison (result is +1, 0 or -1)

      Compare absolute value of MP numbers I   CMPA (MA, MB)         MPCMPA
      Compare MP number with integer       I   COMP (MA, IB)         MPCMPI
      Compare MP number with real          I   COMP (MA, RB)         MPCMPR
      Compare MP number with double-prec.  I   COMP (MA, DB)         MPCMPD
      Compare MP number with rat. IB/IC    I   COMP (MA, IB, IC)     MPCMPQ
      Compare two MP numbers               I   COMP (MA, MB)         MPCOMP

  Relational

      Compare MP* numbers for equality     L   GA .EQ. GB            MPEQ
        (Similarly for .GE.,               L   GA .GE. GB            MPGE
        .GT., .LE., .LT. and .NE.)             etc.                  etc.

  Test

      Three-way branch (n1, n2, n3 labels)     IF (MA) n1,n2,n3      MPSIGA
```

```
              Description                      Res Invocation            Ref

  Field functions

     Sign of MP* number                         I   SGN (GA)              MPSIGA
                                                                          MPSIGB
     Exponent of MP* number                     I   EXPON (GA)            MPEXPA
                                                                          MPEXPB
     IB-th digit of MP number                   I   DIGIT (MA, IB)        MPDGA
                                                                          MPDGB
     Number of MP digits (T)                    I   NUMDIG (GDUMMY)       MPDIGA
                                                                          MPDIGB
     Maximum exponent of MP numbers (M)         I   MAXEXP (MDUMMY)       MPMEXA
                                                                          MPMEXB
     Multiple-precision base (B)                I   BASE (MDUMMY)         MPBASA
                                                                          MPBASB
     Parameter in COMMON /MPCOM/,               I   PARAM (HA)            MPPARA
        argument HA may be any of 'BASE',                                 MPPARB
        'NUMDIG', 'MAXEXP', 'LUN', 'SPTR',
        'MXSPTR', 'MNSPTR', 'MXEXPN',
        'MNEXPN', 'RNDRL', 'KTUNFL',
        'MXUNFL', 'DECPL', 'MT2', 'MXINT',
        'EXWID', 'INRECL', 'INBASE',
        'OUTBAS', 'EXPCH', 'CHWORD',
        or 'ONESCP'

     See under 'Input and output'               M   FIO (IA)              MPFIN
                                                                          MPFOUT
     See under 'Input and output'               M   UNFIO (IA)            MPUNFR
                                                                          MPUNFW

  Miscellaneous

     Unary minus of MP number                   M   -MA                   MPNEG
     Unary minus of packed MP number            P   -PA                   MPNEG
     Assignment of MP number (result = MA)      M   MA                    MPSTR
     Assignment of packed MP number             P   PA                    MPSTR
     Absolute value of MP number                M   ABS (MA)              MPABS
     Absolute value of packed MP number         P   ABS (PA)              MPABS
     Fractional part of MP number               M   FRAC (MA)             MPCMF
     Integer part of MP number                  M   INT (MA)              MPCMIM
     Ceiling function of MP number              M   CEIL (MA)             MPCEIL
     Floor function of MP number                M   FLOOR (MA)            MPFLOR
     Returns true if MA is exact integer        L   INTEGR (MA)           MPINTG
     Maximum of two MP numbers                  M   MAX (MA, MB)          MPMAX
     Maximum of two packed MP numbers           M   MAX (PA, PB)          MPMAX
     Minimum of two MP numbers                  M   MIN (MA, MB)          MPMIN
     Minimum of two packed MP numbers           M   MIN (PA, PB)          MPMIN
     Max (0, MA - MB) for MP numbers            M   DIM (MA, MB)          MPDIM
     Max (0, PA - PB) for packed numbers        M   DIM (PA, PB)          MPDIM
     MA - int(MA/MB)*MB for MP numbers          M   MOD (MA, MB)          MPMOD
     PA - int(PA/PB)*PB for packed numbers      M   MOD (PA, PB)          MPMOD
     abs(MA)*sign(MB) for MP numbers            M   SIGN (MA, MB)         MPSIGN
     abs(PA)*sign(PB) for packed numbers        M   SIGN (PA, PB)         MPSIGN
     Initialization of MP package                   INITIALIZE MP         MPINIT
```

5.3 Synonyms

In some cases Augment will recognize synonyms for the names given in the
column headed 'Invocation' in the summary above.  For example, AMAX1 may
be  used  in  place  of MAX.  A list of synonyms follows.  In most cases
these are usable only for unpacked multiple-precision arguments.  It  is
easy  to  modify or expand the list by making appropriate changes to the
Augment description deck.

```
     Generic name           Synonyms

        ABS                 DABS
        AINT                DINT
        ASIN                ARCSIN, ARSIN, DARSIN, DASIN
        ATAN                ARCTAN, ARTAN, DARTAN, DATAN
        ATAN2               ARTAN2, DATAN2
        COS                 DCOS
        COSH                DCOSH
        DAW                 DDAW
        DIM                 DDIM
        EI                  DEI
        ERF                 DERF
        ERFC                DERFC
        EXP                 DEXP
        GAM                 DGAM, DGAMMA, GAMMA,
                            GAMQ (rational argument only)
        INT                 IDINT, IFIX
        LI                  DLI
        LOG                 ALOG, DLOG, LN, LNI (integer argument only)
        LNGM                ALNGM, DLNGM
        LOG10               ALOG10, DLOG10
        MAX                 AMAX1, DMAX1
        MIN                 AMIN1, DMIN1
        SIGN                DSIGN
        MOD                 AMOD, DMOD
        SIN                 DSIN
        SINH                DSINH
        SQRT                DSQRT
        TAN                 DTAN
        TANH                DTANH
```

5.4 Reserved words

When writing programs which use MP via the Augment interface, it is best
to  avoid  using  the  following  identifiers except with their reserved
meanings as indicated above -

        INITIALIZE,
        MPxxxx (for any letters or digits xxxx),
        MULTIPAK,
        MULTIPLE.

Care  should  also be taken to avoid using the names given in the column
headed 'Invocation' in the summary above, or in  the  list  of  synonyms
above, in a manner which would confuse Augment.

## 6. DESCRIPTION OF MP SUBROUTINES

We give first the method of calling each multiple-precision routine directly, second (third, ...) alternative methods (if any) using the Augment interface described in Sections 4 and 5.

Unless otherwise noted, X, Y and Z denote integer arrays representing multiple-precision variables (often called (unpacked) multiple-precision numbers, or (unpacked) multiple-precision variables, or variables of type MULTIPLE), ERR and LV denote LOGICAL variables, I, J, K, L, IX etc. denote single-precision INTEGER variables, RX, RY etc. denote single-precision REAL variables, DX, DY etc. denote DOUBLE PRECISION variables, and A denotes a packed Hollerith string.

For definitions of B, T, M, LUN, MXR, RNDRL, INBASE, OUTBAS, ... see Section 3.1. 'Space' means the size of the work area in COMMON, i.e. MXR+1-MNSPTR (see Section 3.1). MPGAM, MPLNGM and MPZETA require $O(T^{**}2)$ space, other MP routines require $O(T)$ space.

Time and space bounds such as $O(T^{**}2)$ are as T tends to infinity with everything else fixed. For the definition of the function $m(T)$ appearing in some time bounds, see the description of MPMUL below.

MPABS

CALL MPABS (X, Y)   or   Y = ABS (X)

Sets Y = abs(X) for multiple-precision variables X and Y. X and Y may be both packed or both unpacked. This is an exception to the general rule that an unpacked result is returned even if the argument(s) are packed. The result is exact.

MPADD

CALL MPADD (X, Y, Z)   or   Z = X + Y

Adds X and Y, forming result in Z, where X, Y and Z are multiple-precision numbers.

Rounding is defined by the parameter RNDRL in COMMON /MPCOM/ (see Section 3.1) as follows -

    RNDRL = 0 - Truncate towards zero if X*Y >= 0, away from zero if
                X*Y < 0, in both cases using one guard digit, so the
                result is exact if severe cancellation occurs.

    RNDRL = 1, 2 or 3 - See description of MPCQM. Sufficient guard
                digits are used to give the best possible result.

MPADDI

CALL MPADDI (X, IY, Z)   or   Z = X + IY

Adds multiple-precision X to integer IY giving multiple-precision Z. Rounding options are implemented as for MPADD.

MPADDQ

CALL MPADDQ (X, I, J, Y) or Y = ADDQ (X, I, J)

Adds the rational number I/J to the multiple-precision number X, giving a multiple-precision result in Y. For the effect of rounding options see the descriptions of MPCQM and MPADD.

MPADD3

Called by MPADD, does inner loops of addition. Not recommended for independent use.

MPART1

CALL MPART1 (N, Y) or Y = ART1 (N)

Computes multiple-precision Y = arctan(1/N), for integer N > 1. Uses the Taylor series arctan(x) = x - x**3/3 + x**5/5 - ... , where x = 1/N. Rounding options are implemented as for MPEXP.

MPASIN

CALL MPASIN (X, Y) or Y = ASIN (X)

Returns Y = arcsin(X), for multiple-precision numbers X and Y, where abs(X) <= 1. The result is in the range -pi/2 to +pi/2. Uses MPATAN, so time = O(m(T)T). Rounding options are not yet implemented, no guard digits used.

MPATAN

CALL MPATAN (X, Y) or Y = ATAN (X)

Returns Y = arctan(X) for (packed or unpacked) multiple-precision X, unpacked multiple-precision Y, using an O(m(T)T) method. An asymptotically faster method would be to combine Newton's method with MPTAN. For another asymptotically faster method, see - R. P. Brent, 'Fast multiple-precision evaluation of elementary functions', J. ACM 23 (1976), 242-251, and the description of MPPIGL. Result is in the range -pi/2 to +pi/2. Rounding options are implemented as for MPEXP.

MPATN2

CALL MPATN2 (X, Y, Z) or Z = ATAN2 (X, Y)

Returns Z = arctan (X/Y) if Y nonzero,
        Z = pi/2         if Y = 0,

where X and Y are (both packed or both unpacked) multiple-precision variables, Z is an unpacked multiple-precision variable. Rounding options are implemented as for MPEXP.

MPBASA

I = MPBASA (X) or I = BASE (X)

Returns the multiple-precision base (first word of COMMON /MPCOM/ - see Section 3.1). X is a dummy multiple-precision argument (Augment expects one), I is an integer.

MPBASB

CALL MPBASB (I, X) or BASE (X) = I

Sets the multiple-precision base (first word of COMMON /MPCOM/) to I. I should be an integer such that I > 1 and (8*I*I-1) is representable as a single-precision integer (see Section 3.1). X is a dummy multiple-precision argument (Augment expects one).

Setting the base does not automatically convert multiple-precision numbers to the new base. This can be done by converting to decimal using MPFOUT, changing the base, and converting back using MPFIN.

MPBERN

CALL MPBERN (N, P, X) or X(1) = BERN (N, (PARAM ('MT2') + 1)/2)

Computes the Bernoulli numbers B(2) = 1/6, B(4) = -1/30, B(6) = 1/42, B(8)= -1/30, ... , B(2N), which are defined by the generating function y/(exp(y)-1). N and P are single-precision integers, with 2*P > T+1.

For use without Augment, X should be a one-dimensional integer array of dimension at least P*N. The Bernoulli numbers B(2), B(4), ... , B(2N) are returned in packed format in X, with B(2J) in locations X((J-1)*P+1), ... , X(J*P). Thus, to get B(2J) in unpacked multiple-precision format in Y, one should CALL MPUNPK (X(IX), Y) after calling MPBERN, where IX = (J-1)*P+1.

Alternatively (simpler but nonstandard) - X may be a two-dimensional integer array declared with dimension (P, N1), where N1 >= N and 2*P > T+1. Then B(2), B(4), ... , B(2N) are returned in packed format in X, with B(2J) in X(1,J), ... , X(P,J). Thus, to get B(2J) in the usual multiple-precision format in Y one should CALL MPUNPK (X(1, J), Y) after calling MPBERN.

For use with Augment, declare

        MULTIPAK X(N1)

where N1 >= N, and use (PARAM('MT2') + 1)/2 as the second argument.

The well-known recurrence is unstable (losing about 2J bits of relative accuracy in the computed B(2J)), so we use a different recurrence derived by multiplying the generating function y/(exp(y)-1) by sinh(y/2) and equating coefficients. (This method was suggested by Christian Reinsch, and is faster than the method used in earlier versions of MPBERN.)

The relation

        B(2J) = -2((-1)**J)factorial(2J)zeta(2J)/((2pi)**(2J))

is used if zeta(2J) equals 1 to working accuracy.  The relative error in
B(2J) is $O((J^{2})*(B^{1-T}))$.  For further details, see - R. P. Brent,
'Unrestricted algorithms for elementary and special functions',
Information Processing 80, North Holland, 1980, 613-619.

Time = $O(T*min(N,T)^{2} + N*m(T))$.

Rounding options are not implemented, no guard digits used.

The above remarks assume that N is positive.  If N is negative, they
apply with N replaced by abs(N), except that the precision decreases
linearly, from full precision for B(2) to low precision for B(2N).  This
is faster, and sufficiently accurate for most applications where the
Bernoulli numbers are required to generate coefficients in an
Euler-Maclaurin sum formula.

MPBESJ

CALL MPBESJ (X, NU, Y) or Y = BESJ (X, NU)

Returns Y = J (NU, X), the first-kind Bessel function of order NU, for
small integer NU, multiple-precision X and Y.  The method used is
Hankel's asymptotic expansion if abs(X) is large, the power series if
abs(X) is small, and the backward recurrence method otherwise.  Results
for negative arguments are defined by

     J (-NU, X) = J (NU, -X) = ((-1)**NU) * J (NU, X).

Time = $O(m(T)T)$ for fixed X and NU, increases as X and NU increase,
unless X is large enough for asymptotic series to be used.  Space
required is large (compared to most multiple-precision routines), but
$O(T)$.

Rounding options are not yet implemented.  Uses truncated arithmetic
with some guard digits.

MPBES2

Uses the backward recurrence method to evaluate J (NU, X), where X is a
multiple-precision variable, NU (the index) is an integer, and J is the
Bessel function of the first kind.  Assumes that NU >= 0 and X > 0.  For
normalization the identity

     J(0,X) + 2*J(2,X) + 2*J(4,X) + ... = 1

is used.  Called by MPBESJ and not recommended for independent use.
Rounding options are not implemented, no guard digits used.

MPCAM

CALL MPCAM (A, X) or X = CTM (A) or X = CAM (A)

Converts the (packed) Hollerith string A to a multiple-precision number X. A can be a string of digits acceptable to routine MPIN and terminated by '$', e.g. '-5.367$' or '1E-99$', or one of the following special strings

      'EPS'   (multiple-precision machine-precision, see MPEPS),
      'EUL'   (Euler's constant 0.5772..., see MPEUL),
      'MAXR'  (largest valid multiple-precision number, see MPMAXR),
      'MINR'  (smallest positive multiple-precision number, see MPMINR),
      'PI'    (pi = 3.14..., see MPPI).

Actually, only the first two characters of these special strings are significant. For example, 'MA', 'MAX' and 'MAXR' are equivalent.

The error message '*** MXR TOO SMALL ... ***' is usually caused by omission of the sentinel '$'.

For the effect or rounding options, see the descriptions of MPIN, MPEPS, MPMAXR, MPMINR and MPPI.

Warning to Augment users - use CAM(A) and not CTM(A) if A is declared as an integer array. (Otherwise Augment will generate a call to MPCIM instead of MPCAM.)

MPCDM

CALL MPCDM (DX, Z) or Z = DX or Z = CTM (DX)

Converts double-precision DX to multiple-precision Z. Some numbers will not convert exactly on machines with base other than two, four or sixteen, or if B is not a power of two, or if T is too small. Thus, MPCDM should be used only to obtain starting approximations etc., and not where full multiple-precision accuracy is required. Rounding options are not implemented. For accurate initialization of multiple-precision variables, use MPCAM, MPCIM, MPCQM or MPQPWR.

MPCDM, MPCMD, MPCMDE and MPCMPD are not called by any other routines in the MP package, so they may be omitted if double-precision is not available.

MPCEIL

CALL MPCEIL (X, Y) or Y = CEIL (X)

Sets Y = ceiling (X), i.e. the smallest integer not less than X, where X and Y are unpacked multiple-precision variables. Rounding is defined by RNDRL in COMMON /MPCOM/, as for MPEXP (this is only relevant if X is large and positive, for otherwise ceiling (X) is exactly representable as a multiple-precision number).

MPCHEB

CALL MPCHEB (C, NC, N, IND) or CALL MPCHEB (C, PARAM ('MT2'), N, IND)

Converts the power series coefficients C(1), ... , C(N) to Chebyshev series coefficients (which overwrite C(1), ... , C(N)). It is assumed that on summation of the Chebyshev series the constant term is halved.

IND may have the value 0, -1 or +1, with the following meanings -

  0 - C(j) is the coefficient of x**(j-1) on input, and of T(j-1, x) on output, where T(0, x), T(1, x), ... are the Chebyshev polynomials, e.g.
        T(3, x) = 4*x**3 - 3*x.

 -1 - C(j) is the coefficient of x**(2j-1) on input, and of T(2j-1, x) on output (useful for approximation of odd functions).

 +1 - C(j) is the coefficient of x**(2j-2) on input, and of T(2j-2, x) on output (useful for approximation of even functions).

Conceptually, C is an array of multiple-precision variables. If MPCHEB is called directly, C is a two-dimensional INTEGER array (with first dimension NC at least T+2, second dimension at least N). If Augment is used, C should be of type MULTIPLE, with dimension at least N, and NC should be PARAM ('MT2').

Rounding options are not implemented. Uses truncated arithmetic and no guard digits, time = O(N*N*T), space = O(T).

MPCHEV

CALL MPCHEV (C, NC, N, IND, X, Y)

                    or CALL MPCHEV (C, PARAM ('MT2'), N, IND, X, Y)

Returns Y = Chebyshev series evaluated at X, where X and Y are unpacked multiple-precision variables, C is an array containing the Chebyshev coefficients. For a description of the arguments C, NC, N and IND, see the notes on MPCHEB above.

The algorithm used is described in Chapter 8 of - Modern Computing Methods (Notes on Applied Science No. 16, HMSO, London, 1961).
TIme = O(N*m(T)), space = O(T).

Rounding options are not implemented. Uses truncated arithmetic and no guard digits.

MPCHGB

J = MPCHGB (B1, B2, N)

Returns J such that B1**abs(J) >= B2**abs(N), i.e.

      abs(J) >= abs(N)*log(B2)/log(B1),

and sign(J) = sign(N), where B1, B2 and J are integers, B1 > 1, B2 > 0, B1*B2 <= MXINT.  The value of J returned is close to minimal.  Uses only integer arithmetic.

MPCHK

CALL MPCHK

Checks legality of parameters in COMMON /MPCOM/ (see Section 3.1), and updates MXSPTR.  If an illegal parameter in COMMON /MPCOM/ is detected, an error message is written on unit LUN, and execution is terminated.

MPCIM

CALL MPCIM (I, Z) or Z = I or Z = CTM (I)

Converts single-precision integer I to multiple-precision Z, where abs(I) <= B**T.

MPCIS

CALL MPCIS (X, C, S, BOTH)

Returns  C = cos (X) and S = sin (X)   if BOTH = .TRUE.,
or just  C = cos (X)                   if BOTH = .FALSE.

X, C and S are unpacked multiple-precision variables, BOTH is a logical variable.  Faster than calling MPCOS and MPSIN with the same first argument.  The algorithm is similar to that used in MPEXP (with imaginary argument), time = O(sqrt(T)m(T)).  Rounding options are implemented as follows -

     RNDRL = 0 or 1 - Absolute error < 0.6*B**(-T)
                (but relative error may be large if abs(X) > 1).
     RNDRL = 2 - Lower bound on the true result.
     RNDRL = 3 - Upper bound on the true result.

MPCMD

CALL MPCMD (X, DZ) or DZ = X or DZ = CTD (X)

Converts multiple-precision X to double-precision DZ.  Assumes X is in allowable range for double-precision numbers (otherwise double-precision floating-point overflow or underflow may occur).  There may be some loss of accuracy if B is not a power of the base used for double-precision floating-point arithmetic, or if T is too small.  Rounding options are not implemented.  See also the description of MPCDM.

MPCMDE

CALL MPCMDE (X, N, DX)

Returns integer N and double-precision DX such that (multiple-precision)

     X = DX*OUTBAS**N (approximately),

where 1 <= abs(DX) < OUTBAS (unless X = 0). The default value of OUTBAS is ten (see Section 3.1). It is assumed that X is not so large or small that N overflows. X may be packed or unpacked. Rounding options are not implemented. See also the description of MPCDM.

MPCMEF

CALL MPCMEF (X, N, Y)

Given multiple-precision X, returns integer N and multiple-precision Y such that

    X = (OUTBAS**N)*Y and 1 <= abs(Y) < OUTBAS

(unless X = 0, when N = Y = 0). The default value of OUTBAS is ten (see Section 3.1). It is assumed that X is not so large or small that N overflows.

Rounding options are not fully implemented, but the directed rounding options (RNDRL = 2 or 3) give correct (though not best possible) lower and upper bounds on the true result.

MPCMF

CALL MPCMF (X, Y) or Y = CMF (X) or Y = FRAC (X)

For multiple-precision X and Y, returns fractional part of X in Y, i.e.

    Y = X - integer part of X (truncated towards 0).

Rounding options are irrelevant as the result is exact.

MPCMI

CALL MPCMI (X, I) or I = X or I = CTI (X)

Converts multiple-precision X to integer I, assuming that X is not too large (else use MPCMIM). X is truncated towards zero (irrespective of RNDRL). I is returned as zero if abs(int(X)) > MXINT. The user may check for this possibility by testing if

    ((X(1).NE.0) .AND. (X(2).GT.0) .AND. (I.EQ.0))

is true on return from MPCMI.

MPCMIM

CALL MPCMIM (X, Y) or Y = CMIM (X) or Y = INT (X)

Returns Y = integer part of X (truncated towards 0), for multiple-precision X and Y. Use if abs(Y) > MXINT (otherwise MPCMI is preferable.) X may be packed or unpacked if MPCMIM is called directly.

MPCMP

J = MPCMP (X, Y)

Compares the unpacked multiple-precision numbers X and Y, returning

```
+1 if X > Y,
 0 if X = Y,
-1 if X < Y.
```

X and Y must be unpacked.  If they are packed, MPCOMP should be used instead.  The result is exact.

MPCMPA

J = MPCMPA (X, Y) or J = CMPA (X, Y)

Compares abs(X) with abs(Y) for multiple-precision X and Y, returning

```
+1 if abs(X) > abs(Y),
 0 if abs(X) = abs(Y),
-1 if abs(X) < abs(Y).
```

X and/or Y may be packed or unpacked if MPCMPA is called  directly,  but must be unpacked if called via Augment.  The result is exact.

MPCMPD

J = MPCMPD (X, DI) or J = COMP (X, DI)

Compares multiple-precision X with double-precision DI, returning

```
+1 if X > DI,
 0 if X = DI,
-1 if X < DI.
```

X may be packed or unpacked if MPCMPD is called directly.  Uses MPCDM to convert DI to multiple-precision, so the  result  may  be  incorrect  if MPCDM is inaccurate.  See the description of MPCDM.

MPCMPI

J = MPCMPI (X, I) or J = COMP (X, I)

Compares multiple-precision number X with integer I, returning

```
+1 if X > I,
 0 if X = I,
-1 if X < I.
```

X  may  be  packed or unpacked if MPCMPI is called directly, but must be unpacked if called via Augment.  The result is exact.

MPCMPQ

K = MPCMPQ (X, I, J)  or  K = COMP (X, I, J)

Compares multiple-precision X with the rational number I/J, where I and J are single-precision integers, J nonzero.  Returns

      +1 if X > I/J,
       0 if X = I/J,
      -1 if X < I/J.

Rounding options are irrelevant as the result is exact (the rational number I/J is never computed).

MPCMPR

J = MPCMPR (X, RI)  or  J = COMP (X, RI)

Compares multiple-precision number X with real number RI, returning

      +1 if X > RI,
       0 if X = RI,
      -1 if X < RI.

X may be packed or unpacked if MPCMPR is called directly.  Uses MPCRM to convert RI to multiple-precision, so the result may be incorrect if MPCRM is inaccurate.  See the description of MPCRM.

MPCMR

CALL MPCMR (X, RZ)  or  RZ = X  or  RZ = CTR (X)

Converts multiple-precision X to single-precision real RZ.  Assumes that X is in the allowable range for single-precision real variables (otherwise floating-point overflow or underflow may occur).  There may be some loss of accuracy if B is not a power of the base used for single-precision real arithmetic.  Rounding options are not implemented.

MPCMRE

CALL MPCMRE (X, N, RX)

Returns integer N and single-precision real RX such that multiple-precision

      X = RX*OUTBAS**N (approximately),

where 1 <= abs(RX) < OUTBAS (unless X = 0).  It is assumed that X not so large or small that N overflows.  The default value of OUTBAS is ten (see Section 3.1).  Rounding options are not implemented.

MPCMUL

CALL MPCMUL (XR, XI, YR, YI, ZR, ZI)

Computes the complex product

        ZR + i*ZI = (XR + i*XI) * (YR + i*YI),

where i**2 = -1.  XR, XI, YR, YI, ZR, ZI are unpacked multiple-precision variables.  Rounding options are not implemented.  Uses truncated arithmetic with no guard digits.  Time = $O(m(T))$.

MPCOMP

J = MPCOMP (X, Y) or J = COMP (X, Y)

Compares the multiple-precision numbers X and Y, returning

        +1 if X > Y,
         0 if X = Y,
        -1 if X < Y.

X and/or Y may be packed or unpacked if MPCOMP is called directly, but must be unpacked if called via Augment.  The result is exact.

MPCOS

CALL MPCOS (X, Y) or Y = COS (X)

Returns Y = cos(X) for multiple-precision X and Y, using MPCIS.  X may be packed or unpacked, Y is unpacked.  Time = $O(sqrt(T)m(T))$.  Rounding options are implemented as for MPCIS, so the absolute error is small but the relative error may be large.

MPCOSH

CALL MPCOSH (X, Y) or Y = COSH (X)

Returns multiple-precision Y = COSH(X) for multiple-precision X (not too large).  X may be packed or unpacked, Y is unpacked.  Rounding options are not yet implemented, uses no guard digits.
Time = $O(sqrt(T)m(T))$.

MPCQM

CALL MPCQM (I, J, Q) or Q = CQM (I, J) or Q = CTM (I, J)

Converts the rational number I/J to multiple-precision Q. Rounding options are implemented. Rounding is controlled by the parameter RNDRL in COMMON /MPCOM/ (see Section 3.1), as follows -

  0 - Round towards zero (i.e. chop or truncate). The resulting error is
      less than 1 ulp (unit in the last place).  This is the default
      option if MPINIT or MPSET2 is used.  It is the fastest option, and
      is satisfactory for most purposes.

   1 - Round to nearest, preferring even last digit if a tie occurs.
       Error is less than or equal to 0.5 ulp.  This option gives
       worst-case error bounds half as large as for RNDRL = 0, and
       average-case error considerably better than for RNDRL = 0.
       However, it is expensive in time and space.  If RNDRL = 0 does not
       give sufficient accuracy, it is usually better to increase T than
       to use RNDRL = 1.

   2 - Round down (towards -infinity). Error is less than 1 ulp.  This
       option is useful for interval arithmetic.

   3 - Round up (towards +infinity).  Error is less than 1 ulp.  This
       option is useful for interval arithmetic.

MPCRM

CALL MPCRM (RX, Z) or Z = RX or Z = CTM (RX)

Converts single-precision real RX to multiple-precision Z.  Some numbers
will not convert exactly on machines with base other than two, four or
sixteen, or if B is not a power of 2, or if T is too small.  Thus, MPCRM
should only be used to generate starting approximations etc., and not
where full multiple-precision accuracy is required.  For accurate
initialization of multiple-precision variables, use MPCAM, MPCIM, MPCQM
or MPQPWR.  Rounding options are not implemented.

MPDAW

CALL MPDAW (X, Y) or Y = DAW (X)

Returns

      Y = Dawson's integral of multiple-precision argument X

        = exp(-X**2)*(integral from 0 to X of exp(u**2)du),

where X and Y are multiple-precision variables.  X may be packed or
unpacked if MPDAW is called directly.  Rounding options are implemented
as for MPEXP.

MPDGA

I = MPDGA (X, N) or I = DIGIT (X, N)

Returns the N-th digit of the multiple-precision number X, 0 < N <= T.
Returns zero if X is zero.

MPDGB

CALL MPDGB (I, X, N) or DIGIT (X, N) = I

Sets the N-th digit of the multiple-precision number X to I.  N must be
in the range 0 < N <= T, I must be in the range 0 <= I < B (and I > 0 if
N = 1).  The sign and exponent of X are not changed.

MPDIGA

I = MPDIGA (X) or I = NUMDIG (X)

Returns T, the number of multiple-precision digits (the second word of COMMON /MPCOM/). X is a dummy multiple-precision argument (Augment expects one).

MPDIGB

CALL MPDIGB (I, X) or NUMDIG (X) = I

Sets T, the number of multiple-precision digits (second word of COMMON /MPCOM/), to I, which should be an integer in the range 1 < I < MT2-1. X is a dummy multiple-precision argument (Augment expects one).

MPDIGS

J = MPDIGS (N)

Returns the number of multiple-precision (base B) digits required for the equivalent of at least N floating 'decimal' places. The inequality

    B**(MPDIGS - 1) >= OUTBAS**(N - 1)

will be satisfied, usually with the minimal such value of MPDIGS. If OUTBAS has its default value (ten), the working precision may be set to the equivalent of at least N floating decimal places by the statement

       PARAM ('NUMDIG') = MPDIGS (N)        (using Augment)
or
       CALL MPPARB (MPDIGS (N), 'NUMDIG')   (not using Augment).

MPDIGV

J = MPDIGV (C)

Returns the integer value

       0 , ... , 9 , 10, ... , 15 if C is the corresponding character
       '0', ... , '9', 'A', ... , 'F'

and the value returned is less than INBASE (default value ten, see Section 3.1). Otherwise returns -1. The character value should be represented in an integer variable as if read under A1 format.

MPDIGW

C = MPDIGW (J)

Returns the character '0', ... , 'F' if J is the corresponding integer,
                     0 , ... , 15,

returns '*' otherwise. The character is stored in an integer as if read under A1 format.

MPDIM

CALL MPDIM (X, Y, Z) or Z = DIM (X, Y)

Sets Z = X - min (X, Y) = max (0, X-Y) for multiple-precision variables X, Y and Z. X and Y may be both packed or both unpacked, Z is unpacked. Rounding options are implemented as for MPSUB.

MPDIV

CALL MPDIV (X, Y, Z) or Z = X/Y

Sets Z = X/Y, for multiple-precision X, Y and Z. Y must not be zero. Uses MPDIVL for small T or if RNDRL > 0, MPREC and MPMUL otherwise. Thus time = O(m(T) + RNDRL*T**2). Rounding options are implemented as for MPCQM.

MPDIVI

CALL MPDIVI (X, IY, Z) or Z = X/IY

Divides unpacked multiple-precision X by the nonzero single-precision integer IY, giving multiple-precision Z in time O(T). This is much faster than division by a multiple-precision number. Rounding options are implemented as for MPCQM.

MPDIVL

CALL MPDIVL (X, Y, Z)

Sets Z = X/Y for unpacked multiple-precision variables X, Y and Z (Y nonzero). Uses the 'schoolboy' or 'long division' algorithm, so time = O(T**2). An alternative method is to use MPREC and MPMUL. This would be faster for large T, but MPDIVL is faster for small T. Rounding options are implemented as for MPCQM.

MPDIV2

Routine called by MPDIVI, not recommended for independent use.

MPDIV3

Routine called by MPDIVL, not recommended for independent use.

MPDUMP

CALL MPDUMP (X)

Dumps the (packed or unpacked) multiple-precision number X on unit LUN. Writes the sign, exponent and digits in suitable I format (or just the sign if X is zero). Embedded blanks should be interpreted as zeros. MPDUMP is sometimes useful for debugging, but is not recommended in general for output of MP numbers.

MPEI

CALL MPEI (X, Y) or Y = EI (X)

Returns

        Y = ei (X) = -E1 (X)
          = (principal value integral from -infinity to X of exp(u)/u du),

for multiple-precision numbers X and Y, using the power series for small abs(X), the asymptotic series for large abs (X), and the continued fraction for medium negative X.  The relative error in Y is small unless X is very close to the zero 0.37250741078136663446... of ei(X), and then the absolute error in Y is $O(B^{**}(1-T))$.  In any case the error in Y could be induced by an $O(B^{**}(1-T))$ relative perturbation in X.  Rounding options are not yet implemented.  Time = $O(m(T)T)$.

MPEPS

CALL MPEPS (X) or X = CTM ('EPS')

Sets multiple-precision X to the (multiple-precision) machine precision, that is an upper bound on the smallest representable positive number X such that the relative error in the basic multiple-precision operations of addition, subtraction, multiplication and division does not exceed X unless the result underflows.  In fact

        X = 1.01*B**(1-T)   (rounded up)   if RNDRL = 0,
          =  0.5*B**(1-T)   (rounded up)   if RNDRL = 1,
          =      B**(1-T)                  if RNDRL = 2 or 3.

MPEQ

LV = MPEQ (X, Y) or LV = (X .EQ. Y) or IF (X .EQ. Y) ...

Returns logical value LV of (X .EQ. Y) for multiple-precision X and Y. MPEQ must be declared LOGICAL unless the Augment interface is used. X and/or Y may be packed or unpacked.

MPERF

CALL MPERF (X, Y) or Y = ERF (X)

Returns Y = erf(X)
          = sqrt(4/pi)*(integral from 0 to X of exp(-u**2) du)

for multiple-precision X and Y.  X may be packed or unpacked if MPERF is called directly, must be unpacked if MPERF is called via Augment. Rounding options are implemented as for MPEXP.

MPERFC

CALL MPERFC (X, Y) or Y = ERFC (X)

Returns Y = erfc(X) = 1 - erf(X) for multiple-precision numbers X and Y. X may be packed or unpacked if MPERFC is called directly, must be unpacked if MPERFC is called via Augment. Rounding as for MPEXP.

MPERF2

Computes exp(X**2)*(integral from 0 to X of exp(-u*u) du)

for multiple-precision X, using the power series for small X, and MPEXP for large X. Called by MPERF, not recommended for independent use. Rounding options are not implemented, no guard digits are used.

MPERF3

Trys to return

    exp(X**2)*(integral from X to infinity of exp(-u**2)du),

or

    exp(-X**2)*(integral from 0 to X of exp(u**2) du),

using the asymptotic expansion, where X is a multiple-precision variable. The condition for success is approximately that

    X > sqrt(T*ln(B)).

Called by MPERF, MPERFC and MPDAW, not recommended for independent use. Rounding options are not implemented, uses no guard digits.

MPERR

CALL MPERR

This routine is called when a fatal error condition is encountered, and after a message has been written on logical unit LUN. As supplied in the multiple-precision package, MPERR writes a message which includes the version number in the format YYMMDD, then stops. On many systems it can easily be modified to give a trace-back (e.g. with Univac 1100 Fortran V this can be obtained by replacing STOP by RETURN 0).

Since MPERR is only called when a fatal error condition has been detected, it does not return to the calling routine.

MPERRM

CALL MPERRM (A)

The argument A is a packed Hollerith string terminated by '$'. MPERRM writes the string on unit LUN, then calls MPERR. Only the first 71 characters of A are significant. They will be preceeded and followed by '***'. MPERRM is called by MP routines when a fatal error is encountered, and does not return to the calling routine.

MPEUL

CALL MPEUL (G) or G = CTM ('EUL')

Returns multiple-precision G = Euler's constant (gamma = 0.57721566...) to almost full multiple-precision accuracy. The method was discovered by Edwin McMillan and Richard Brent, and is faster than the method of Sweeney (used in earlier versions of MPEUL). See - R. P. Brent and E. M. McMillan, 'Some new algorithms for high-precision computation of Euler's constant', Math. Comp. 34(1980), 305-312. Time = $O(T^{**}2)$. Rounding options are implemented as for MPEXP.

MPEXP

CALL MPEXP (X, Y) or Y = EXP (X)

Returns Y = exp(X) for multiple-precision X and Y. X may be packed or unpacked if MPEXP is called directly. The method used is described in - R. P. Brent, 'The complexity of multiple-precision arithmetic', in The Complexity of Computational Problem Solving, Queensland University Press, Brisbane, 1976, 126-165. See also - R. P. Brent, 'Unrestricted algorithms for elementary and special functions', Information Processing 80, North-Holland, 1980, 613-619. Time = $O(sqrt(T)m(T))$.

Rounding is controlled by the parameter RNDRL in COMMON /MPCOM/ (see Section 3.1), as follows -

    0 or 1 - error less than 0.6 ulp (units in the last place),

    2      - computed result is lower bound on the true result,

    3      - computed result is upper bound on the true result.

MPEXPA

I = MPEXPA (X) or I = EXPON (X)

Returns the exponent of the multiple-precision number X (or large negative exponent -M if X is zero). X may be packed or unpacked if MPEXPA is called directly.

MPEXPB

CALL MPEXPB (I, X) or EXPON (X) = I

Sets the exponent of the multiple-precision number X to I, unless X is zero (when its exponent is unchanged). X must be a valid multiple-precision number (either zero or normalized). X may be packed or unpacked if MPEXPB is called directly. Note that multiple-precision underflow/overflow will occur if I is too small/large.

MPEXP1

CALL MPEXP1 (X, Y) or Y = EXP1 (X)

Returns

    Y = exp(X) - 1

where X and Y are multiple-precision numbers and -1 < X < 1. Uses an O(sqrt(T)m(T)) algorithm described in - R. P. Brent, 'The complexity of multiple-precision arithmetic', in Complexity of Computational Problem Solving, Univ. of Queensland Press, Brisbane, 1976, 126-165. Rounding options are implemented as for MPEXP.

MPFIN

CALL MPFIN (LUNIT, X) or X = FIO (LUNIT)

Reads (unpacked) multiple-precision number X from Fortran logical unit LUNIT (LUNIT >= 0). It is assumed that a record of length INRECL <= 80 may be read in '(80A1)' format from unit LUNIT, and then converted to multiple-precision using MPIN. Thus, MPFIN may be used to read free-format fixed or floating-point numbers, one per record, from unit LUNIT. The input base INBASE has default value ten (i.e. decimal input - see Section 3.1). For further details see under MPIN.

MPFLOR

CALL MPFLOR (X, Y) or Y = FLOOR (X)

Sets Y = floor (X), i.e. the largest integer not exceeding X, where X and Y are unpacked multiple-precision variables. Rounding is defined by RNDRL in COMMON /MPCOM/, as for MPEXP (this is only relevant if X is large and negative, for otherwise floor (X) is exactly representable as a multiple-precision number).

MPFOUT

CALL MPFOUT (X, N) or FIO (N) = X

Writes the multiple-precision number X on unit LUN in floating-point format, with N significant 'decimal' digits (N > 1). The exponent field width is EXWID (default is 6, including exponent character EXPCH and sign - see Section 3.1). The default exponent character EXPCH is 'E' (or '$' if OUTBAS = 15 or 16), and the default output base OUTBAS is ten (i.e. decimal output). Rounding options are implemented as for MPOUTE. X may be packed or unpacked if MPFOUT is called directly, but must be unpacked if called via Augment.

MPGAM

CALL MPGAM (X, Y) or Y = GAM (X)

Computes multiple-precision Y = Gamma(X) for multiple-precision argument X, using MPGAMQ if 240*X is a small integer, otherwise using MPLNGM. Space required is about the same as for MPLNGM, i.e. O(T**2), time = O(T**3). Rounding options are not yet implemented and no guard digits are used.

MPGAMQ

CALL MPGAMQ (I, J, X) or Y = GAMQ (I, J) or Y = GAM (I, J)

Returns

    X = Gamma (I/J),

where X is multiple-precision, I and J are small integers. The method used is reduction of the argument to (0, 1) and then a direct expansion of the defining integral truncated at a sufficiently high limit, using 2T digits to compensate for cancellation. Time = O(T**2) if I/J is not too large. If I/J > 100 (approximately) it is faster to use MPGAM. (MPGAMQ is very slow if I/J is is very large, because the relation Gamma(X+1) = X*Gamma(X) is used repeatedly.) Rounding options are not yet implemented.

MPGCD

CALL MPGCD (K, L)

Returns K = K/GCD and L = L/GCD, where GCD is the greatest common divisor of the initial K and L (single-precision integers). Called by various MP routines.

MPGCDA

CALL MPGCDA (X, Y, Z) or Z = GCD (X, Y)

Returns Z = greatest common divisor of X and Y. (By definition, GCD (X, 0) = GCD (0, X) = abs(X), GCD (X, Y) >= 0.) X, Y and Z are integers represented as multiple-precision numbers, and must satisfy abs(X) < B**T, abs(Y) < B**T. The method is a straight-forward implementation of the Euclidean algorithm, and Lehmer's trick is not used, although it might be faster (see Amer. Math. Monthly 45 (1938), 227-233). Time = O(T**2). X and/or Y may be packed or unpacked if MPGCDA is called directly. Rounding options are irrelevant as the result is exact.

MPGCDB

CALL MPGCDB (X, Y)

Returns (X, Y) as (X/Z, Y/Z) where Z is the GCD of initial X and Y, which are integers represented as multiple-precision numbers, and must satisfy abs(X) < B**T, abs(Y) < B**T. Time = O(T**2). Rounding options are irrelevant as the result is exact.

MPGD

J = MPGD (N)

Returns ceiling (ln (max (1, abs(N))) / ln(B)),

i.e. the minimum J >= 0 such that

    B**J >= abs(N).

This function is useful for computing the number of guard digits required for various (multiple-precision) base B calculations.

MPGD3

CALL MPGD3 (N, TG)

Sets T = T + 1 + MPGD (100*N) if it is safe to call MPGD, and the same or slightly more otherwise. Also sets TG to the new value of T. N and TG are integer arguments. Called by various MP routines when it is necessary to increase the working precision.

MPGE

LV = MPGE (X, Y) or LV = (X .GE. Y) or IF (X .GE. Y) ...

Returns the logical value of (X > Y) for multiple-precision X and Y. MPGE must be declared LOGICAL unless the Augment interface is used. X and/or Y may be packed or unpacked.

MPGET

J = MPGET (X, N)

Returns X(N), where X is an integer array of dimension at least N. Necessary to avoid some compiler diagnostics.

MPGT

LV = MPGT (X, Y) or LV = (X .GT. Y) or IF (X .GT. Y) ...

Returns the logical value of (X > Y) for multiple-precision X and Y. MPGT must be declared LOGICAL unless the Augment interface is used. X and/or Y may be packed or unpacked.

MPHANK

Tries to compute the Bessel function J (NU, X) using Hankel's asymptotic expansion, for nonnegative integer NU and multiple-precision X. Time = O(T**3). Called by MPBESJ, not recommended for independent use. Rounding options are not implemented, uses no guard digits.

MPIMUL

CALL MPIMUL (IY, X, Z) or Z = IY*X

Equivalent to

        CALL MPMULI (X, IY, Z) or Z = X*IY.

See the description of MPMULI for further details.

## MPIN

CALL MPIN (C, X, N, ERROR)

Converts the characters (assumed to have been read under NA1 format) in C(1), ... , C(N) to a multiple-precision number in X. If C represents a valid number, ERROR is returned as .FALSE. If C does not represent a valid number, ERROR is returned as .TRUE. and X as zero. Leading and trailing blanks are allowed, embedded blanks (except between the number and its sign) are forbidden. If there is no decimal point one is assumed to lie just to the right of the last decimal digit. X is a multiple-precision number, C an integer array, N an integer, and ERROR logical.

The input base is determined by the parameter INBASE in COMMON /MPCOM/ (see Section 3.1). INBASE has default value ten, but may be set to any value in two, ... , sixteen.

```
     Examples of valid numbers     Examples of invalid numbers

      -   123456789                12 345
          3.14159                  123.456E -67
         -44.                      1.2.3
         .0001234                  E123
          123.456D789              64.4E+
         -.1234566-789             ++12.3
         +999+88                   E3.
```

Rounding is determined by the parameter RNDRL in COMMON /MPCOM/ as follows -

    0 or 1 - Round to (approximately) the nearest representable base B, T-digit number. The error is less than 0.6 units in the last (base B) place.

    2 - Round down (towards -infinity) to a base B, T-digit number,

    3 - Round up (towards +infinity) to a base B, T-digit number.

## MPINE

CALL MPINE (C, X, N, J, ERROR)

MPINE is the same as MPIN except that the result (X) is multiplied by INBASE**J, where J is a single-precision integer. For details of the other arguments, see MPIN. Useful for floating-point input of multiple-precision numbers. The user can read the exponent into J (using any suitable format) and the fraction into C (using NA1 format), then call MPINE to convert to multiple-precision.

Note - in early versions of the MP package, MPIN was unable to deal with numbers with exponents (e.g. 5.3E-26). Now that MPIN can deal with such numbers there is no real need for MPINE, so it is included only for compatibility with early versions of the package.

MPINF

CALL MPINF (X, N, UNIT, IFORM, ERR)  or  ERR = MPINF (X, N, UNIT, IFORM)

or IF (MPINF (X, N, UNIT, IFORM)) ...

Reads N words from logical unit abs(UNIT) using format IFORM, then converts to multiple-precision number X using routine MPIN. IFORM should contain a format which allows for reading N words in A1 format, e.g. '(80A1)'.

ERR is returned as .TRUE. if MPIN could not interpret input as a multiple-precision number or if N not positive, otherwise .FALSE. In the former case X is returned as zero.

Note - for a more convenient, though less flexible, method of reading multiple-precision numbers, see MPFIN.

MPINIT

CALL MPINIT (MP) or INITIALIZE MP

Declares blank COMMON (see Section 3.1) and calls MPSET2 to initialize parameters. MP is a dummy integer argument. The Augment declaration

    INITIALIZE MP

causes the statement

    CALL MPINIT (MP)

to be generated.

Warning - as distributed MPINIT assumes output unit LUN = 6, at most 25 multiple-precision digits (i.e. MT2 = 27), MNSPTR = 1, and MXR = 500. If the Augment description deck is changed, MPINIT should be changed accordingly (see Section 9.2).

MPINTG

LV = MPINTG (X) or LV = INTEGR (X) or IF (INTEGR (X)) ...

Returns .TRUE. if the (unpacked) multiple-precision number X is an exact integer, .FALSE. otherwise. LV is a LOGICAL variable. MPINTG must be declared LOGICAL if it is called directly.

MPIO

CALL MPIO (C, N, UNIT, IFORM, ERR)

Writes C(1), ... , C(N) in format IFORM if UNIT > 0,
Reads  C(1), ... , C(N) in format IFORM if UNIT <= 0,

in both cases using logical unit abs(UNIT). Unformatted I/O is performed if IFORM = 'U'. C is an integer array of dimension at least N.

N and UNIT are integers, ERR is a logical variable. ERR is returned as .TRUE. iff N <= 0.  (It would be desirable to return ERR = .TRUE. if an I/O error occurred.  This is easily done on most systems, but can not be done with ANSI Standard Fortran. See Section 9.3.)

MPIS

LV = MPIS (J, K)

Assumes that J and K are INTEGER words each containing one character in the leftmost character position (read under A1 format or set by a 1Hx data statement or unpacked by MPUPK).  MPIS returns .TRUE. if the two characters are equal, .FALSE. otherwise.  LV is a LOGICAL variable, and MPIS must be declared to be LOGICAL.

On most machines, MPIS (J, K) is equivalent to (J .EQ. K), but on the Burroughs B6700 it is equivalent to (J .IS. K).  For conversion notes, see Section 9.2.

MPKADD

CALL MPKADD (X, Y, Z) or Z = X + Y

The same as CALL MPADD (X, Y, Z) except that X and Y are packed multiple-precision variables.  (They may be packed or unpacked if MPKADD is called directly - similarly for MPKDIV, MPKDVI, MPKIML, MPKMLI, MPKMUL and MPKSUB.)  For further details, see under MPADD.

MPKDIV

CALL MPKDIV (X, Y, Z) or Z = X/Y

The same as CALL MPDIV (X, Y, Z) except that X and Y are packed multiple-precision variables.  For further details see under MPDIV.

MPKDVI

CALL MPKDVI (X, IY, Z) or Z = X/IY

The same as CALL MPDIVI (X, IY, Z) except that X is a packed multiple-precision variable.  For further details, see under MPDIV.

MPKIML

CALL MPKIML (IY, X, Z) or Z = IY*X

The same as CALL MPMULI (X, IY, Z) except that X is a packed multiple-precision variable.  For further details, see under MPMULI.

MPKMLI

CALL MPKMLI (X, IY, Z) or Z = X*IY

The same as CALL MPMULI (X, IY, Z) except that X is a packed multiple-precision variable.  For further details, see under MPMULI.

MPKMUL

CALL MPKMUL (X, Y, Z) or Z = X*Y

The same as CALL MPMUL (X, Y, Z) except that X and Y are packed multiple-precision variables. For further details, see under MPMUL.

MPKSUB

CALL MPKSUB (X, Y, Z) or Z = X - Y

The same as CALL MPSUB (X, Y, Z) except that X and Y are packed multiple-precision variables. For further details, see under MPSUB.

MPK3V

CALL MPK3V (MPXX, X, Y, Z)

Calls MPXX (XU, YU, Z) after unpacking X and Y to give XU and YU. X and Y are packed or unpacked multiple-precision variables, Z is an unpacked multiple-precision variable, MPXX is a subroutine taking three unpacked multiple-precision arguments and returning a result as the third argument (e.g. MPMUL). Called by MPKADD, MPKMUL, MPKSUB etc.

MPK3V2

CALL MPK3V2 (MPXX, X, N, Y)

Calls MPXX (XU, N, Z) after unpacking X to give XU. X is a packed or unpacked multiple-precision variable, N is an integer, Z is an unpacked multiple-precision variable, MPXX is a subroutine taking three arguments of the correct types and returning a result as the third argument (e.g. MPMULI). Called by MPKDVI, MPKMLI etc.

MPLARG

CALL MPLARG (MXINT)

Returns MXINT <= maximum representable INTEGER of the form 2**k - 1. Integer arithmetic must be performed exactly on integers in the range -MXINT, ... , +MXINT, so on some machines (e.g. Burroughs B6700 and Cyber 76) MXINT must be smaller than the largest representable integer. We say that the 'effective' wordlength is k+1 bits.

Note - The version of MPLARG supplied with the MP package does not work on all machines. For conversion notes, see Section 9.2.

MPLE

LV = MPLE (X, Y) or LV = (X .LE. Y) or IF (X .LE. Y) ...

Returns logical value of (X < Y) for multiple-precision X and Y. MPLE must be declared type logical unless Augment interface used. X and/or Y may be packed or unpacked.

MPLG10

CALL MPLG10 (X, Y) or Y = LOG10 (X)

Returns Y = ln(X)/ln(10) = log(X) to base 10, for multiple-precision X and Y. X may be packed or unpacked, Y is unpacked. Uses MPLN and MPLNI. For restrictions, time bounds, and the effect of rounding options, see the description of MPLN.

MPLI

CALL MPLI (X, Y) or Y = LI (X)

Returns

>      Y = Li(X) = logarithmic integral of X
>        = principal value integral from 0 to X of du/ln(u),

using MPEI. X and Y are multiple-precision numbers, X > 0, X .NE. 1. X may be packed or unpacked if MPLI is called directly. Rounding options are not yet implemented. Error in Y could be induced by an $O(B^{**}(1-T))$ relative perturbation in X followed by similar perturbation in Y. Thus relative error in Y is small unless X is close to 1 or to the zero 1.45136923488338105028... of Li(X). Time = $O(m(T)T)$.

MPLN

CALL MPLN (X, Y) or Y = LN (X) or Y = LOG (X) or Y = ALOG (X)

Returns Y = ln(X), for multiple-precision X and Y, using MPLNS. X may be packed or unpacked, Y is unpacked. The integer part of ln(X) must be representable as a single-precision integer. Time = $O(sqrt(T)m(T))$. For small integer X, MPLNI is faster. Asymptotically faster methods exist (e.g. the Gauss-Salamin method), but are not useful unless T is large (see the descriptions of MPATAN, MPEXP1, MPLNGS and MPPIGL). Rounding options are implemented as for MPEXP.

MPLNGM

CALL MPLNGM (X, Y) or Y = LNGM (X)

Returns multiple-precision Y = ln(Gamma(X)) for positive multiple-precision X, using Stirling's asymptotic expansion. (X may be packed or unpacked if MPLNGM is called directly.) Slower than MPGAMQ (unless X large) and uses more space, so use MPGAMQ and MPLN if X is rational and not too large, say X less than 100. Time = $O(T^{**}3)$.

Space = $nl*((T+3)/2) + O(T)$ words, where nl is the number of terms used in the asymptotic expansion, $nl < 2+0.125*T*ln(B)$. Thus space = $O(T^{**}2)$ words.

Rounding options are not yet implemented, uses no guard digits.

MPLNGS

CALL MPLNGS (X, Y)

Returns Y = ln(X) for multiple-precision X and Y, using the Gauss-Salamin algorithm based on the arithmetic-geometric mean iteration. This is described in - R. P. Brent, 'Multiple-precision zero-finding methods and the complexity of elementary function evaluation', in Analytic Computational Complexity (ed. by J.F. Traub), Academic Press, 1976, 151-176. (If abs(X-1) < 1/B, MPLNS is used.) Time = O(ln(T)m(T)) + O(T**2) if abs(X-1) > 1/B, as for MPLNS otherwise.

MPLNGS is slower than MPLN unless T is large (greater than about 500), so it is recommended for testing purposes only. Rounding options are implemented as for MPEXP.

MPLNI

CALL MPLNI (N, X) or X = LNI (N) or X = ALOG (N)
              or X = LN  (N) or X = LOG  (N)

Returns multiple-precision X = ln(N) for small positive integer N, using a rapidly converging series and MPL235. Time = O(T**2), faster than MPLN. Rounding options are implemented as for MPEXP.

MPLNS

CALL MPLNS (X, Y) or Y = LNS (X)

Returns multiple-precision Y = ln(1+X) if X is a multiple-precision number satisfying the condition abs(X) < 1/B, error otherwise. X may be packed or unpacked if MPLNS is called directly. Uses Newton's method to solve the equation exp1(-z) = X, then sets Y = -z. (Here exp1(z) = exp(z) - 1 is computed using MPEXP1.) Time = O(sqrt(T)m(T)). Rounding options are implemented as for MPEXP.

MPLT

LV = MPLT (X, Y) or LV = (X .LT. Y) or IF (X .LT. Y) ...

Returns logical value of (X < Y) for multiple-precision X and Y. MPLT must be declared LOGICAL unless the Augment interface used. X and/or Y may be packed or unpacked.

MPL235

Computes ln ((2**i)*(3**j)*(5**k)) = i*ln(2) + j*ln(3) + k*ln(5) for small integer i, j and k. ln(81/80), ln(25/24) and ln(16/15) are calculated first, using rapidly converging series. Time = O(T**2). Called by MPLNI, not recommended for independent use. Rounding options are not implemented, uses no guard digits.

MPMAX

CALL MPMAX (X, Y, Z) or Z = MAX (X, Y)

Sets Z = max (X, Y) where X, Y and Z are multiple-precision variables. X and/or Y may be packed or unpacked if MPMAX is called directly. Rounding options are irrelevant as the result is exact.

MPMAXR

CALL MPMAXR (X) or X = CTM ('MAXR')

Sets X to the largest possible positive multiple-precision number, i.e. $B^{**}M - B^{**}(M-T)$. Rounding options are irrelevant as the result is exact.

MPMEXA

I = MPMEXA (X) or I = MAXEXP (X)

Returns the maximum allowable exponent of multiple-precision numbers (the third word of COMMON /MPCOM/ - see Section 3.1). X is a dummy multiple-precision argument (Augment expects one).

MPMEXB

CALL MPMEXB (I, X) or MAXEXP (X) = I

Sets the maximum allowable exponent of multiple-precision numbers to I. I should be greater than T, and 4*I should be representable as a single-precision integer. X is a dummy multiple-precision argument (Augment expects one).

MPMIN

CALL MPMIN (X, Y, Z) or Z = MIN (X, Y)

Sets Z = min (X, Y) where X, Y and Z are multiple-precision variables. X and/or Y may be packed or unpacked if MPMIN is called directly. Rounding options are irrelevant as the result is exact.

MPMINR

CALL MPMINR (X) or X = CTM ('MINR')

Sets X to the smallest positive normalized multiple-precision number, i.e. $X = B^{**}(-M)$, where M is the third word of COMMON /MPCOM/ (see Section 3.1). Rounding options are irrelevant as the result is exact.

MPMLP

Performs inner multiplication loop for MPMUL. Carries are not propagated in inner loop, which saves time at the expense of space. This routine is usually called more often than any other MP routine, so it is worthwhile to optimize it.

MPMOD

CALL MPMOD (X, Y, Z) or Z = MOD (X, Y)

Returns  Z = X - Y*int(X/Y)  for multiple-precision X, Y and Z.  X and Y
may be both packed or both unpacked, Z is unpacked.  Here  'int'  means
the integer part, truncated towards zero.

Notes - 1. MPMOD returns Z as zero if Y is zero.
        2. Time = O (max (1, ln(abs(X/Y))) * T**2)),
           which is large if abs(X) >> abs(Y).
        3. Rounding options are irrelevant as the result is exact.

MPMOVE

CALL MPMOVE (X, TX, Y, TY)

Assumes  that  X  and  Y  are  multiple-precision numbers with TX and TY
digits respectively.  X may be packed or unpacked, Y  is  unpacked  (but
not  necessarily  initialized).   MPMOVE  moves  X  to  Y,  padding with
trailing zeros if TX < TY, or rounding as specified by RNDRL  (see  the
description  of  MPCQM)  if TX > TY.  Note that TX and TY are integer
arguments, and the  value  of  T  (second  word  of  COMMON  /MPCOM/)  is
irrelevant.

MPSTR (X, Y) is equivalent and faster if TX = TY = T and X is  unpacked.

MPMUL

CALL MPMUL (X, Y, Z) or Z = X*Y

Multiplies  X  and  Y,  returning  result  in  Z, for multiple-precision
variables X, Y and Z.  The simple O(T**2) algorithm  is  used,  with  at
least  two  guard  digits.   Advantage is taken of any zero digits in X, but
not in Y.  Asymptotically faster algorithms are known (see  for  example
Knuth,  Vol.  2),  but  are  difficult  to  implement  in Fortran in an
efficient and machine-independent manner.

In description of other multiple-precision routines, m(T) is the time to
perform T-digit multiple-precision multiplication.  Thus m(T) = O(T**2)
with  the present version of MPMUL, but m(T) = O(T.log(T)log(log(T))) is
theoretically possible (see Knuth, Vol. 2).

Rounding  options  are  implemented.   They are controlled by the parameter
RNDRL in COMMON /MPCOM/ (see Section 3.1) as follows -

  RNDRL = 0 - truncate towards zero, error less than 1.01 units  in  the
              last place (ulp),

  RNDRL = 1 - round to nearest representable multiple-precision  number,
              preferring last digit even in case of a tie, error at most
              0.5 ulp (unless the result underflows),

   RNDRL = 2 - round down (towards -infinity), error less than 1 ulp,

   RNDRL = 3 - round up (towards +infinity), error less than 1 ulp.

Note that RNDRL = 0 is the fastest (approximately twice as fast as the other options) because less guard digits are required. Thus, RNDRL = 0 is recommended for general use and is the default option.

MPMULI

CALL MPMULI (X, IY, Z) or Z = X*IY

Multiplies multiple-precision X by single-precision integer IY giving multiple-precision Z. Time = O(T), which is faster than MPMUL. Multiplication by 1 may be used to normalize a number even if the last digit is B. Rounding options are implemented as for MPCQM.

MPMULQ

CALL MPMULQ (X, I, J, Y) or Y = MULQ (X, I, J)

Multiplies multiple-precision X by rational number I/J, giving multiple-precision Y. Here I and J are single-precision integers, J nonzero. Time = O(T).

Rounding options are determined by RNDRL in COMMON /MPCOM/ (see Section 3.1) as follows -

   RNDRL = 0 - as for MPMULI followed by MPDIVI,

   RNDRL = 1, 2, or 3 - as for MPCQM.

Augment users should note that Y = MULQ (X, I, J) is usually faster than Y = X * CTM (I,J).

MPMULS

CALL MPMULS (X, I, J, K, L)

Sets X = X*I*J/(K*L) for unpacked multiple-precision X, integer I, J, K and L (K*L nonzero). Calls MPMULQ once if I*J and K*L are not too large, otherwise calls MPMULQ twice. Rounding is not best possible, but the directed rounding options give correct upper and lower bounds.

MPNE

LV = MPNE (X, Y) or LV = (X .NE. Y) or IF (X .NE. Y) ...

Returns logical value of (X .NE. Y) for multiple-precision X and Y. MPNE must be declared LOGICAL unless the Augment interface used. X and/or Y may be packed or unpacked.

MPNEG

CALL MPNEG (X, Y) or Y = -X

Sets Y = -X for multiple-precision numbers X and Y.  X and Y may be both packed or both unpacked. (This violates the usual rule of returning an unpacked result regardless of whether the argument is packed or unpacked.  The reason is that Augment does not allow unary operators to return a type different from that of the variable which they operate on.)

MPNEW

CALL MPNEW (I)

Returns integer index I such that words I, ... , I+T+1 of blank COMMON are available for use.  Sets SPTR = I+T+2 and updates MXSPTR (see Section 3.1).  MPSTOV is called if MXR is too small.

Note - CALL MPNEW (I) is equivalent to CALL MPNEW2 (I, T+2).

MPNEW2

CALL MPNEW2 (I, J)

Returns integer index I such that words I, ... , I+abs(J)-1 of blank COMMON are available for use.  Sets SPTR = I+abs(J) and updates MXSPTR (see Section 3.1).  MPSTOV is called if MXR is too small.

MPNZR

Normalizes and rounds a multiple-precision number (represented in a nonstandard format, with separate sign and exponent, and possibly some guard digits).  Rounding depends on RNDRL in COMMON /MPCOM/ - see the description of MPCQM.  Called by MPADD, MPDIVI, MPMUL, MPMULI, etc., and not recommended for independent use.

MPOUT

CALL MPOUT (X, C, P, N)

Converts multiple-precision X to FP.N format in C, which may be printed under PA1 format. (Here FP.N means sign, P-N-2 'decimal' places before the 'decimal' point, and N places after it.)  Note that N = -1 is allowed, and effectively gives IP format (i.e., sign and P-1 'decimal' digits).  Digits after the 'decimal' point are blanked out if they could not be significant.  The output base is OUTBAS in COMMON /MPCOM/ (see Section 3.1), in the range two to sixteen, with default value ten.  P and N are integers, C is an integer array of dimension at least P.  X may be packed or unpacked.

Rounding is defined by the parameter RNDRL in COMMON /MPCOM/, as follows -

   0 or 1 -  Round to approximately the nearest decimal number representable in the specified format. The error in the conversion to decimal is less than 0.6 units in the last decimal place.

   2 -       Round down (towards -infinity) to a decimal number representable in the specified format.

   3 -       Round up (towards +infinity) to a decimal number representable in the specified format.

Efficiency is higher if B is a power of 10 than if not. Time = O(T**2).

Note - MPFOUT and MP40D are more convenient to use, although less flexible than MPOUT.

MPOUTE

CALL MPOUTE (X, C, J, P)

Assumes X is a multiple-precision number and C an integer array of dimension at least P > 3. J and P are integers. On return J is the exponent (to base OUTBAS) of X and the fraction is in C, ready to be printed in A1 format. For example, we could print J and C in format '(I10, 1X, PA1)'. The fraction has one place before the 'decimal' point and P-3 after. The output base is OUTBAS in COMMON /MPCOM/ (see Section 3.1), with default value ten. Rounding is not the best possible, but the directed rounding options (RNDRL = 2 and 3) give correct lower and upper bounds on the true result (see the description of MPOUT). X may be packed or unpacked.

Note - MPFOUT is a more convenient (though less flexible) output routine.

MPOUTF

CALL MPOUTF (X, P, N, IFORM, ERR)  or  ERR = MPOUTF (X, P, N, IFORM)
                                   or  IF (MPOUTF (X, P, N, IFORM)) ...

Writes multiple-precision number X on logical unit LUN (fourth word of COMMON /MPCOM/) in format IFORM after converting to FP.N 'decimal' representation using MPOUT. The output base is OUTBAS, with default value ten. For further details see description of MPOUT. IFORM should contain format which allows for output of P words in A1 format, plus any desired spacing etc. ERR is returned as .TRUE. iff P <= 0 (see description of MPIO). X may be packed or unpacked if MPFOUT is called directly. For rounding options see MPOUT. Time = O(T**2).

Note - a more convenient (though less flexible) output routine is MPFOUT.

MPOUT2

CALL MPOUT2 (X, C, P, N, NB)

Same as MPOUT except that output representation is in base NB, where two < NB < sixteen, e.g. NB = eight gives octal output, NB = sixteen gives hexadecimal. Output digits are 0123456789ABCDEF. X is a multiple-precision number, P, N and NB are integers, C is an integer array of dimension at least P. Time = $O(T^{**}2)$. For rounding options see the description of MPOUT.

Note - MPOUT2 is superfluous now that OUTBAS determines the output base for MPOUT. It is included for compatibility with earlier versions of the MP package.

MPOVFL

CALL MPOVFL (X)

Called on multiple-precision overflow, that is when the exponent of the multiple-precision number X would exceed M. Execution is terminated with an error message after calling MPMAXR(X) and setting MXEXPN = M+1 (see Section 3.1).

MPPACK

CALL MPPACK (X, Y) or Y = X or Y = CTP (X)

Assumes that X is a multiple-precision number represented as usual in an integer array of dimension at least T+2, and Y is an integer array of dimension at least int((T+3)/2). X is stored in a compact format in Y, and may be retrieved by calling MPUNPK (Y, X).

MPPACK and MPUNPK are useful if space is critical, for example when working with large arrays of multiple-precision numbers. X may be packed or unpacked if MPPACK is called directly. (If X is packed the effect of MPPACK is the same as that of MPSTR.)

Augment users - X is type MULTIPLE, Y is type MULTIPAK.

The packed format is as follows -

    word 2 = exponent (as for unpacked multiple-precision numbers),

    words 1 and 3, ... , int((T+3)/2) = base B digits packed two per word (digits pq are represented as p*B+q), with sign attached to word 1.

Thus, zero is represented in both packed and unpacked formats by a zero first word, with the following words undefined. The first word has absolute value greater than 1 if and only if the multiple-precision number is nonzero and is represented in the packed format.

MPPARA

I = MPPARA (A) or I = PARAM (A)

Returns the integer value of the word of COMMON /MPCOM/ corresponding to the Hollerith string A. For the allowable values of A, see the description of MPPARM.

MPPARB

CALL MPPARB (I, A) or PARAM (A) = I

Sets the word of COMMON /MPCOM/ corresponding to the Hollerith string A to the integer value I. For the allowable values of A, see the description of MPPARM. For the allowable values of I, see Section 3.1.

MPPARC

CALL MPPARC (N, J)

Sets the N-th word of COMMON /MPCOM/ to the integer value J, for 0 < N < 24. Faster than MPPARB, but less mnemonic.

MPPARM

CALL MPPARM (HOLL, SET, J)

HOLL is a Hollerith string, SET is a Boolean variable, and J is an integer. The first three characters of HOLL should agree with the first three characters of one of the following -

   (1)  'BASE'       (2)  'NUMDIG'    (3)  'MAXEXP'    (4)  'LUN'
   (5)  'MXR'        (6)  'SPTR'      (7)  'MXSPTR'    (8)  'MNSPTR'
   (9)  'MXEXPN'    (10)  'MNEXPN'   (11)  'RNDRL'    (12)  'KTUNFL'
  (13)  'MXUNFL'    (14)  'DECPL'    (15)  'MT2'      (16)  'MXINT'
  (17)  'EXWID'     (18)  'INRECL'   (19)  'INBASE'   (20)  'OUTBAS'
  (21)  'EXPCH'     (22)  'CHWORD'   (23)  'ONESCP'

The corresponding word in COMMON /MPCOM/ will be set to the integer J if SET = .TRUE., or its value will be returned in J if SET = .FALSE. For the meanings and default settings of these parameters in COMMON /MPCOM/, see Section 3.1.

MPPARN

J = MPPARN (N)

Returns the value of the N-th word of COMMON /MPCOM/, for 0 < N < 24, zero otherwise. Faster than MPPARA, but less mnemonic.

MPPI

CALL MPPI (X) or X = CTM ('PI')

Sets multiple-precision X = pi = 3.14159... to the available precision. Uses the formula

    pi = 16*arctan(1/5) - 4*arctan(1/239)

of Machin. Time = $O(T^{**}2)$. (For asymptotically faster methods, see the description of MPPIGL.) Rounding options are implemented as for MPEXP.

MPPIGL

CALL MPPIGL (X)

Sets multiple-precision X = pi = 3.14159... to almost the available precision. Uses the Gauss-Legendre algorithm. This method requires time $O(ln(T)m(T))$, so it is slower than MPPI if $m(T) = O(T^{**}2)$, but would be faster for large T if a faster multiplication algorithm were used (see the description of MPMUL). For a description of the Gauss-Legendre algorithm see – R. P. Brent, 'Multiple-precision zero-finding and the complexity of elementary function evaluation', in Analytic Computational Complexity (edited by J.F. Traub), Academic Press, l976, 151-176.

Rounding options are not implemented, uses no guard digits. Recommended for testing purposes only.

MPPOLY

CALL MPPOLY (X, Y, IC, N)

Sets Y = IC(1) + IC(2)*X + ... + IC(N)*X**(N-1), where X and Y are multiple-precision numbers, and IC is an integer array of dimension at least N > 0. Rounding is not best possible, but the directed rounding options (RNDRL = 2 and 3) give correct lower and upper bounds on the true result.

MPPWR

CALL MPPWR (X, N, Y) or Y = X**N

Returns Y = X**N, for multiple-precision X and Y, integer N, with 0**0 = 1. X may be packed or unpacked, Y is unpacked. Rounding options are implemented as for MPEXP.

MPPWRA

CALL MPPWRA (X, N, Y)

Returns Y = X**abs(N) for (unpacked) multiple-precision X and Y, integer N, with 0**0 = 1. Uses truncated rather than rounded arithmetic, and no guard digits, but directed rounding options give correct upper and lower bounds. Called by MPEXP, MPPWR etc., not recommended for independent use (use MPPWR instead).

MPPWR2

CALL MPPWR2 (X, Y, Z) or Z = X**Y

Returns Z = X**Y for multiple-precision numbers X, Y and Z, where X is positive (X = 0 is allowed if Y > 0). X and Y may be both packed or both unpacked, Z is unpacked. Uses MPLN and MPEXP, so slower than MPPWR and MPQPWR. Rounding options are implemented as for MPEXP.

MPQPWR

CALL MPQPWR (I, J, K, L, X) or X = QPWR (I, J, K, L)

Sets multiple-precision X = (I/J)**(K/L) for integers I, J, K and L (with the obvious conditions to ensure that a real result is defined). Uses MPROOT if abs(L) is small, otherwise uses MPLNI and MPEXP. Rounding options are implemented as for MPEXP.

Augment users - X = QPWR (I, J, K, L) is usually faster than
              X = CTM (I, J) ** CTM (K, L).

MPREC

CALL MPREC (X, Y) or Y = REC (X)

Returns Y = 1/X, for multiple-precision X and Y. Uses MPDIVL if T is small or RNDRL > 0, MPROOT otherwise. Time = O(m(T)) if RNDRL = 0, O(T**2) if RNDRL > 0.

Augment users - Y = REC (X) is faster than Y = 1/X.

MPRESN

CALL MPRESN (N)

Restores T, M and RNDRL which are assumed to have been saved on the stack in blank COMMON by a call to MPSAVN (N). The integer N should be as returned by MPSAVN. SPTR is restored to its value before the call to MPSAVN (i.e. N). For further details, see the description of MPSAVN.

MPRES2

CALL MPRES2 (N)

Has the same effect as CALL MPRESN (N) except that SPTR is not changed.

MPREVR

CALL MPREVR (I)

If I > 0 and RNDRL = 2 or 3, RNDRL is set to 3 or 2 respectively (this reverses the direction of rounding - see the description of MPCQM).

MPRND

CALL MPRND (X, TX, Y, TY, K)

Moves  X + S*K*abs(X)*B**(-TY)  appropriately rounded to Y, where X is an unpacked multiple-precision number with TX digits, Y is an unpacked multiple-precision number with TY digits, K is an integer, and

        S = 0 if RNDRL = 0 or 1,
           -1 if RNDRL = 2 (rounding down),
           +1 if RNDRL = 3 (rounding up).

Note that RNDRL = 0 has the same effect as RNDRL = 1, and that the value of T is irrelevant.

MPROOT

CALL MPROOT (X, N, Y) or Y = ROOT (X, N)

Returns Y = X**(1/N) for integer N such that a real result is defined, (packed or unpacked) multiple-precision X and (unpacked) multiple-precision Y. Uses Newton's method without divisions, so time = O(m(t)). Rounding options are implemented as for MPEXP.

Augment users - Y = ROOT (X, N) is faster than Y = X ** CTM (1, N)
                  (and Y = X**(1/N) is incorrect as 1/N is an
                    integer expression, hence usually zero).

MPSAVN

CALL MPSAVN (N)

Saves T, M and RNDRL on the stack in blank COMMON (they may be restored by calling MPRESN). Returns N = old value of SPTR.

        T is saved in word N of blank COMMON,
        M is saved in word N+1, and
        RNDRL is saved in word N+2.

Note - N and SPTR must not have the same address.

MPSCAL

CALL MPSCAL (X, N, J)

Sets X = X*(N**J) for unpacked multiple-precision X, integer J, and small positive integer N. Intended for use when 1 < N <= 100. Rounding is not best possible, but the directed rounding options give correct upper and lower bounds.

MPSET

CALL MPSET (LUN, DECPL, MT2, MXR)

Has the same effect as

       CALL MPSET2 (LUN, DECPL, MT2, 1, MXR)

(see the description of MPSET2). MPSET is redundant, but is included for compatibility with earlier versions of the MP package.

MPSETR

CALL MPSETR (N)

Sets the parameter RNDRL in COMMON /MPCOM/ to N. N should be an integer with value 0, 1, 2 or 3 (see Section 3.1).

MPSET2

CALL MPSET2 (LUN, DECPL, MT2, MNSPTR, MXR)

MPSET2 may be called to initialize the parameters in COMMON /MPCOM/. All five arguments are of type INTEGER. LUN is the Fortran logical unit to be used for error messages. The base (B) and number of digits (T) are set to give the equivalent of at least DECPL > 0 floating decimal digits. MT2 should be the dimension of the arrays used for MP variables, so an error occurs if the computed value of T exceeds MT2-2. MXR is the size of blank COMMON declared in the calling program, and MNSPTR is the first location of blank COMMON available for use by MP routines. (They will use words MNSPTR, MNSPTR+1, ... , MXR of blank COMMON (counting from 1) for working storage.)

Let MXINT be the largest integer of the form 2**k - 1 representable as a signed integer on the machine (for a more precise definition, see the description of MPLARG). The computed B and T satisfy the conditions

   B**(T-1) >= 10**(DECPL-1)   and   8*B*B-1 <= MXINT. Approximately minimal T and maximal B satisfying these conditions are chosen.

MPSET2 also sets the following default values. For the meanings of the various parameters, see Section 3.1.

```
     M = int (MXINT/4),  SPTR = MNSPTR,  MXSPTR = MNSPTR,  RNDRL  =  0,
     KTUNFL  = 0,  MXUNFL = 0,  MNEXPN = M+1,  MXEXPN = -M,  MXINT = as
     above,  EXWID = 6,  INRECL = 80,  INBASE = 10,   OUTBAS  =   10,
     EXPCH  =  'E',  CHWORD = value  returned  by  MPUPW,  ONESCP = 0 or
     1 (see Section 3.1).
```

If these are not all as desired, change after the call to MPSET2. For the methods of doing this, see Section 3.2.

MPSIGA

I = MPSIGA (X) or I = SGN (X)

Returns the sign of a packed or unpacked multiple-precision number X, i.e.

```
     +1 if X > 0,
      0 if X = 0,
     -1 if X < 0.
```

MPSIGB

CALL MPSIGB (I, X) or SGN (X) = I

Sets the sign of a (packed or unpacked) multiple-precision number X to the sign of I. (X is set to zero if I = 0.) The exponent and digits of X are unchanged, but the result must be a valid (packed or unpacked) multiple-precision number.

MPSIGN

CALL MPSIGN (X, Y, Z) or Z = SIGN (X, Y)

Sets Z = abs(X)*sign(Y) for (both packed or both unpacked) multiple-precision variables X and Y, unpacked multiple-precision Z, where

```
     sign (Y) = +1 if Y > 0,
              =  0 if Y = 0,
              = -1 if Y < 0.
```

MPSIN

CALL MPSIN (X, Y) or Y = SIN (X)

Returns Y = sin(X) for multiple-precision X and Y. X may be packed or unpacked, Y is unpacked. Uses MPCIS, so time = O(sqrt(T)m(T)).

Rounding options are implemented as for MPCIS. The absolute error in the computed result is small, i.e. $O(B^{**}(-T))$, but the relative error may be large if X is close to a nonzero multiple of pi.

MPSINH

CALL MPSINH (X, Y) or Y = SINH (X)

Returns Y = sinh(X) for multiple-precision numbers X and Y, X not too large.  X may be packed or unpacked, Y is unpacked.  The method is to use MPEXP or MPEXP1, so time = O(sqrt(T)m(T)).  Rounding options are not yet implemented, uses no guard digits.

MPSQRT

CALL MPSQRT (X, Y) or Y = SQRT (X)

Returns Y = sqrt(X) for packed or unpacked multiple-precision X > 0, unpacked multiple-precision Y.  Uses MPROOT, so time = O(m(T)). Rounding options are implemented as for MPEXP.  (RNDRL = 0 or 1 gives the exact result if it is exactly representable.)

MPSTOV

CALL MPSTOV (N)

MPSTOV is called if the working space in blank COMMON is too small.  If possible, the space should be expanded by at least N words, and MXR should be increased by the number of words expanded.  The new space must be contiguous with the old (i.e. it must include words MXR+1, ... , MXR+N of blank COMMON).

As distributed MPSTOV does nothing, because the method of expanding the working space is machine and operating system dependent (see Section 9.3).

MPSTR

CALL MPSTR (X, Y) or Y = X

Sets Y = X for multiple-precision X and Y.  X may be packed or unpacked. Y will be packed if X is packed, unpacked if X is unpacked.  (This is an exception to the general rule that an unpacked result is returned even if the argument(s) are packed.)

MPSUB

CALL MPSUB (X, Y, Z) or Z = X - Y

Computes   Z = X - Y   for multiple-precision X, Y and Z.   Rounding options are implemented - see the description of MPADD with Y replaced by -Y.

MPSUBA

CALL MPSUBA (X, Y, I)

Returns I = max (1, X(1) - Y(1)) for integer arrays X and Y. Called by MPLARG and MPUPW with Hollerith arguments X and Y.

MPTAN

CALL MPTAN (X, Y) or Y = TAN (X)

Sets Y = tan(X) for multiple-precision X and Y. X may be packed or unpacked, Y is unpacked. Uses MPCIS, so time = O(sqrt(T)m(T)). Rounding options are not implemented, but some guard digits are used.

MPTANH

CALL MPTANH (X, Y) or Y = TANH (X)

Returns Y = tanh(X) for multiple-precision X and Y. X may be packed or unpacked, Y is unpacked. Time = O(sqrt(T)m(T)). Rounding options are implemented as for MPEXP.

MPTLB

I = MPTLB (J)

Returns an upper bound on abs(J)*ln(B)/ln(2), using only integer arithmetic. Called by various MP routines.

MPUNFL

CALL MPUNFL (X)

Called on multiple-precision underflow, that is when the exponent of the multiple-precision number X would be less than 1-M. The underflow counter KTUNFL is incremented. Other action depends on the parameters RNDRL and MXUNFL in COMMON /MPCOM/, as follows -

  If RNDRL < 2 then

     if MXUNFL = 0 (the default option), X is set to zero and execution
                 continues,

     if MXUNFL > 0, X is set to zero and execution continues unless
                 MXUNFL underflows have occurred, when execution is
                 terminated by a call to MPERR.

  If RNDRL = 2, action is as above except that X is set to the smallest
                 negative representable number, i.e. -(B**(-M)).

  If RNDRL = 3, action is as above except that X is set to the smallest
                 positive representable number, i.e. B**(-M). (See
                 also under MPMINR.)

MPUNFR

CALL MPUNFR (LUNIT, X) or X = UNFIO (LUNIT)

Reads a multiple-precision number X from Fortran logical unit LUNIT, assuming it has been written using MPUNFW (with the same B and T). X is unpacked.

MPUNFR and MPUNFW are faster than MPFIN and MPFOUT, so should be used if temporary storage of multiple-precision numbers is required.

MPUNFW

CALL MPUNFW (X, LUNIT) or UNFIO (LUNIT) = X

Writes an multiple-precision number X without formatting on Fortran logical unit LUNIT. X may be packed or unpacked if MPUNFW is called directly, but must be unpacked if called via Augment. To save space on LUNIT, X is packed before being written, so an unformatted record of length int((T+3)/2) words is is actually written. Multiple-precision numbers written with MPUNFW may be read with MPUNFR (so long as the values of B and T have not been changed). See also the description of MPUNFR above.

MPUNPK

CALL MPUNPK (Y, X) or X = Y or X = CTM (Y)

Unpacks the multiple-precision number which is stored in packed format in Y, and stores the result in X. This reverses the effect of

    CALL MPPACK (X, Y).

If called directly, Y may be packed or unpacked, and in the latter case the effect is the same as that of MPSTR. For further details see the description of MPPACK.

Augment interface users - X is type MULTIPLE,
                         Y is type MULTIPAK.

MPUPDT

CALL MPUPDT (J)

Sets MXEXPN = max (J, MXEXPN) and MNEXPN = min (J, MNEXPN).

MPUPK

CALL MPUPK (SOURCE, DEST, LDEST, LFIELD)

This subroutine unpacks a packed Hollerith string SOURCE, placing one character per word in the integer array DEST (as if read in A1 format). It continues unpacking until it finds a sentinel ($) or until it has filled LDEST words of the array DEST. The length of the unpacked string is returned in LFIELD. Thus 0 <= LFIELD <= LDEST.

MPUPW

CALL MPUPW (W, C, N)

W and N are INTEGER variables, C is an INTEGER array.  When called with a packed character string in W, MPUPW returns

      N = number of characters per word
and
      C(1), ... , C(N) = the characters in W (unpacked,
                         left justified, blank filled).

Important note

MPUPW is machine-dependent, and the version supplied with the MP package does not work on all machines (e.g. machines which do not perform integer arithmetic on full words, machines in which characters are stored in an unusual order, and machines with decimal or sign-and-magnitude arithmetic).  For conversion hints, see Section 9.2.

MPZETA

CALL MPZETA (N, X) or X = ZETA (N)

Returns multiple-precision X = zeta(N) for integer N > 1, where zeta(N) is the Riemann zeta function (the sum from j = 1 to infinity of j**(-N)).  Uses the Euler-Maclaurin series unless N = 2, 3, 4, 6 or 8. In the worst case space = k*T/2 + O(T), where k is the number of terms used in the Euler-Maclaurin series, k = O(T).  Time = O(T**3) in general, O(T**2) if N = 2, 3, 4, 6 or 8.  Rounding options are implemented as for MPEXP.

MP40D

CALL MP40D (N, X)

Output routine called by TEST program. Writes multiple-precision X to N 'decimal' places on unit LUN.  The output base is OUTBAS (default value ten - see Section 3.1).  It is assumed that -OUTBAS < X < OUTBAS**2. Rounding options are implemented as for MPOUT.

MP40F

CALL MP40F (N, X)

Equivalent to

      CALL MPFOUT (X, N)

MP40F is redundant, but is included for compatibility with earlier versions of the MP package.

# 7. TEST PROGRAMS

## 7.1 TEST

This main program computes the constants given in Appendix A of Knuth, The Art of Computer Programming, Vol. 3. The constants are printed in the same order as they are given in Knuth. Their correct values (rounded to 40 decimal places) are also given in the comments in the source of TEST.

TEST computes the constants to 40 decimal places, but to increase the accuracy it is only necessary to change the statement PLACES = 40, and possibly the parameters of the call to MPSET2 and the dimensions of the arrays, as described in the comments in the source of TEST. The constants are given to 1000 decimal places in - R. P. Brent, 'Knuth's constants to 1000 decimal and 1100 octal places', Tech. Report 47, Computer Centre, Australian National University, Canberra (Sept. 1975).

Execution time (CPU SUPs) on a Univac 1100/82 computer using FTN (level 9R1) is about 2.464 seconds.

## 7.2 TEST2

This main program tests various multiple-precision routines, especially those not called by program TEST. It computes the constants given (in a different order) in Computer Approximations (by Hart, Cheney, Lawson, Maehly, Mesztenyi, Rice, Thacher and Witzgall), John Wiley, 1968, Appendix C, pp. 182-183, and various other constants which are described in the comments in the source of TEST2.

Most of the constants are computed to 40 significant figures, with working precision equivalent to at least 42 significant figures. To increase the precision, it is only necessary to make a few changes described in the comments. The correct output on a Univac 1100/82 is given in the comments. The output may be slightly different on other machines, especially if they have integer wordlength other than 36 bits.

Execution time is about 25 times longer than that of TEST.

## 7.3 Other test programs

The EXAMPLE program (given in Section 2) and the JACOBI program (given in Section 4.3) test various routines in the MP package, although not as thoroughly as TEST and TEST2 do. The JACOBI program also provides a partial test of the Augment precompiler and the MP description deck (see Section 4).

8. COMPARISON WITH EARLIER VERSIONS OF MP

8.1 New routines and capabilities

For a brief description of the differences between MP version 781124 and earlier versions, see Section 1.3.  The main differences between version 800207 (or later) of the MP package and version 781124 are -

  1. Introduction of the labelled common block COMMON /MPCOM/ (see Section 3.1).

  2. Improved dynamic storage allocation.  A stack is now implemented, with stack pointer SPTR, minimum and maximum pointers MNSPTR and MXSPTR (see Section 3.1).  The environment may be saved with MPSAVN, working space allocated with MPNEW or MPNEW2, and the environment later restored with MPRESN.

     Earlier versions of the MP package did not have a genuine stack, and it was necessary for the high-level routines to know how much space the low-level routines required, which made it very difficult to modify any low-level routines.  Now it is much easier to add high-level routines or to modify low-level routines, as it is not necessary to know the space requirements of routines called by them or calling them.

  3. Although the stack is still in blank COMMON (for a reason given in Section 1.2), it no longer must start at the first word of blank COMMON.  Thus, routines which use blank COMMON for other purposes can call MP routines.  The facility to start the stack anywhere in blank COMMON is also useful on systems where working storage can be obtained at runtime.

  4. Rounding options have been implemented, both for the basic arithmetic operations, and for most of the elementary and special functions implemented in the MP package (see Section 6).  Thus, implementation of a multiple-precision interval-arithmetic package should now be straightforward.

  5. Many routines in the MP package now accept both packed and unpacked arguments, and the Augment description deck has been modified accordingly.  To achieve this the packed representation had to be changed (see below).

  6. The MP routines are now independent of the Fortran routines ALOG, SQRT etc., which makes MP useful for testing them.

  7. The MP routines now depend very little on single-precision REAL arithmetic. Only the conversion and comparison routines MPCDM, MPCMD, MPCMDE, MPCMPD, MPCMPR, MPCMR, MPCMRE and MPCRM use REAL or DOUBLE PRECISION arithmetic.  This makes the MP package useful for testing the correctness of REAL arithmetic, and practically eliminates machine-dependence of the results obtained using the package.  It also makes implementation of most of the routines in the MP package possible on small machines which do not have floating-point (i.e. REAL) arithmetic.

The following routines have been added to the MP package since the release of version 781124. For descriptions of these routines, see Section 6.

MPATN2, MPCEIL, MPCHEB, MPCHEV, MPCHGB, MPCIS, MPCMP, MPCMPD, MPCMPQ, MPCMUL, MPDIGS, MPDIGV, MPDIGW, MPDIM, MPDIVL, MPDIV2, MPDIV3, MPERRM, MPFIN, MPFLOR, MPFOUT, MPGD, MPGD3, MPGET, MPIMUL, MPINTG, MPIS, MPKADD, MPKDIV, MPKDVI, MPKIML, MPKMLI, MPKMUL, MPKSUB, MPK3V, MPK3V2, MPLARG, MPLG10, MPMOD, MPMOVE, MPMULS, MPNEW, MPNEW2, MPPARA, MPPARB, MPPARC, MPPARM, MPPARN, MPPWRA, MPRESN, MPRES2, MPREVR, MPRND, MPSAVN, MPSCAL, MPSETR, MPSET2, MPSIGN, MPSTOV, MPSUBA, MPTLB, MPUNFR, MPUNFW, MPUPDT, MPUPW.

8.2 Incompatibilities with earlier versions of MP

The parameters B, T, M, LUN and MXR, which were formerly the first five words of blank COMMON, are now the first five words of the labelled common block COMMON /MPCOM/ (see Section 3.1). Thus, any routines which manipulate these parameters directly will have to declare them in COMMON /MPCOM/.

The format for packed multiple-precision numbers has been changed – the exponent is now stored in the second word, not the first. The reason for this change is that with the new format it is possible to tell, by inspecting the first word of a multiple-precision number, whether it is packed or unpacked (unless it is zero, when only the first word is defined anyway). For more details, see the description of MPPACK in Section 6.

Some routines used to return an INTEGER error flag. The type of all error flags is now LOGICAL. In all cases the integer value 0 has been replaced by the logical value .FALSE., and the integer value 1 has been replaced by the logical value .TRUE.. Routines affected are MPERF3, MPHANK, MPIN, and MPINE.

The calling sequence of MPADD3 has been changed. This routine should not be called directly by users of the MP package.

The routines MPINE, MPOUT2, MPSET and MP40F are now redundant, and may be deleted from later versions of the MP package. It is recommended that MPIN, MPOUT (with the correct setting of OUTBAS), MPSET2 and MPFOUT (respectively) should be used instead.

The following routines have been removed from the MP package since the release of version 781124 -

MPADD2, MPCLR, MPEXT, MPKSTR, MPMUL2, MPSIN1, MP40E, MP40G, TESTV and TIMEMP.

MPKSTR is no longer needed, as MPSTR now works for both packed and unpacked arguments. The test program TESTV may be reconstructed by making trivial changes to the program TEST. The other routines are low-level routines whose disappearance should not affect a user of the MP package.

# 9. INSTALLATION INSTRUCTIONS

## 9.1 Essentials

MP is normally distributed on an unlabelled, 9 track, Ebcdic or Ascii, 800 or 1600 fpi magnetic tape, with logical record length 80 characters, blocking factor 12, and the following 6 files -

       File 1 - Comments and EXAMPLE program.
       File 2 - MP subroutines.
       File 3 - Test programs TEST and TEST2.
       File 4 - This User's Guide (duo-case version).
       File 5 - Augment description deck and JACOBI program using it.
       File 6 - This User's Guide (upper-case version).

To install MP, read these six files. Print file 4 or 6 (the User's Guide) using the first character as standard Fortran printer control.

Check the source of routines MPINIT, MPIS, MPLARG, and MPUPW (in file 2), make any necessary changes, then compile the routines in files 1, 2 and 3 (preferably with compiler optimization options turned off).

Convert compiled routines from file 2 into a relocatable library.

Execute the programs EXAMPLE, TEST and TEST2 from files 1 and 3 (after linking to required routines from file 2) and check that output is correct (see the comments in the source of the programs). Output is on unit 6 unless the first argument in each call to MPSET2 is changed.

If all has gone well, try recompiling file 2 with compiler optimization options turned on, and rerun the test programs. Any problems which appear are probably due to bugs in your compiler, not in the MP routines. (Such problems exist with some releases of the Univac FTN and DEC 10 (F10) compilers, for example.) The routines whose optimization is most worthwhile are MPADD3, MPDIVI, MPDIV2, MPDIV3, MPMLP and MPNZR.

The Augment description deck is supplied with the MP package, but the Augment precompiler is not. If you want to use the Augment interface, obtain Augment from the Programming Services Supervisor, Mathematics Research Center, 610 Walnut Street, Madison, Wisconsin 53706, and get it running. This should not be too hard if you have a Univac 1100, IBM 360/370, CDC 6000/7000, DEC 10, or Honeywell 600/6000, as Augment has already been implemented on these machines. (Augment is written mainly in portable Fortran, but there are a few machine-dependent routines.)

Next, use the description deck supplied in file 5 and run the JACOBI test program which follows the description deck in file 5 (after inserting a few machine-dependent control cards). It will be necessary to change the dimension statements in the description deck and modify routine MPINIT if your machine has wordlength less than 16 bits, and desirable to do likewise if the wordlength is greater than 16 bits (because arrays used for multiple-precision variables will be declared larger than necessary). See the comments in MPINIT regarding declaration of COMMON if your system insists that all common blocks be declared in the main program.

9.2 Conversion notes

To convert MPINIT and the description deck, let

    i = dimension of arrays for MP variables (see the description deck
       and MPINIT),

    j = dimension of arrays for packed MP variables (see the
       description deck),

    k = size of work area in blank COMMON
      = MXR + 1 - MNSPTR  (see Section 3.1).

Suppose precision equivalent to at least d decimal places is required on a machine with effective wordlength w bits (MXINT = $2^{**}(w-1)-1$ must be representable as a signed integer - see also the description of MPLARG in Section 6).

Then
    i = T + 2,

    j = int ((T + 3)/2),
and
    k <= T*T + 15*T + 200   (usually much less),
where
    T = ceiling (1 + (d-1)*log(10)/log(B))

is the number of base B digits to be used, B = $2^{**}int(w/2 - 2)$.

To convert MPIS on a Burroughs 6000/7000 machine, change 'EQ' to 'IS' in the first executable statement of MPIS.  On some other machines it may be necessary to mask off some characters or shift right before the comparison.

To convert MPLARG on a Cyber 76, Burrough 6000/7000, or PDP 11 machine, set the appropriate logical variable in the DATA statements of MPLARG to .TRUE..  On other machines it may be necessary to set WDLEN to the effective wordlength (for integer arithmetic) in MPLARG.

To convert MPUPW on most machines, replace the body of MPUPW by

```
      INTEGER W, C(chword)
      N = chword
      DECODE (N, 10, W) C
   10 FORMAT (80A1)
```

where 'chword' is the number of characters per word.  On some CDC machines the arguments of DECODE must be given in a different order. (If chword > 10, trivial changes are required in MPSET2 and MPUPW.)

## 9.3 Desirable changes

The following machine-dependent changes are desirable but not essential, and may not be possible on some systems.

MPIO could be modified to return ERR = .TRUE. on a read/write error.

MPUPK could be modified to replace a compiler-generated end-of-string sentinel (if any) by '$', so that string arguments of MPCAM need not be terminated by '$'. See the DATA statement for SENTNL in MPUPK.

MPSTOV could be modified to expand the working storage when necessary.

MPERR could be modified to give a trace-back when an error is detected.

MPDIGV could be made much more efficient. This would be worthwhile if MPIN or MPFIN were to be used extensively.

## 9.4 Note on Fortran 77

The MP routines do not conform strictly to the new 'Fortran 77' Standard (ANS X3.9-1978). There are two main violations of the standard. One is that arrays which are actual parameters of subprograms, and are declared with dimension (1), should instead be declared with dimension (*), e.g.

        SUBROUTINE MPABC (X, Y)
        INTEGER X(1), Y(1)
            ...

should be replaced by

        SUBROUTINE MPABC (X, Y)
        INTEGER X(*), Y(*)
            ...

The other violation is the use of the Hollerith data type, which is not included in the Fortran 77 Standard.

Most Fortran 77 compilers accept these violations as 'extensions' to the new standard, in order to be compatible with the old Fortran Standard (ANS X3.9-1966). This is true, for example, of the Univac 1100 FTN compiler (level 9R1). Thus, we do not anticipate any major problems with Fortran 77 compilers.